\documentclass[aps,prd,twocolumn,superscriptaddress,nofootinbib,preprintnumbers]{revtex4-2}
\usepackage{style}
\usepackage[normalem]{ulem}
\usepackage{ragged2e}

\definecolor{mulberry}{rgb}{0.5,0,0.5}
\definecolor{elderberry}{rgb}{0.3,0,0.7}
\definecolor{raspberry}{RGB}{255, 87, 51}
\definecolor{gooseberry}{RGB}{0, 171, 79}
\definecolor{blueberry}{RGB}{75, 134, 247}
\definecolor{winterberry}{RGB}{255, 135, 0}

\usepackage[capitalize]{cleveref}

\crefname{appendix}{appendix}{appendices}
\Crefname{appendix}{Appendix}{Appendices}

\definecolor{DarkSlateNavy}{rgb}{0.106, 0.106, 0.227}
\definecolor{DeepPurple}{rgb}{0.294, 0.180, 0.514}
\definecolor{HypotheticBlue}{rgb}{0.216, 0.494, 0.722}
\definecolor{HypotheticGreen}{rgb}{0.302, 0.686, 0.290}
\definecolor{HypotheticOrange}{rgb}{1.000, 0.498, 0.000}

\newcommand{\eV}{\mathinner{\mathrm{eV}}}
\newcommand{\keV}{\mathinner{\mathrm{keV}}}
\newcommand{\MeV}{\mathinner{\mathrm{MeV}}}
\newcommand{\GeV}{\mathinner{\mathrm{GeV}}}
\newcommand{\TeV}{\mathinner{\mathrm{TeV}}}

\newcommand{\Mpl}{\mathinner{M_\mathrm{pl}}}

\newcommand{\cbrac}[1]{\left( #1 \right) }

\allowdisplaybreaks

\definecolor{RevisionBlue}{RGB}{0,72,158}
\newif\ifrevisioncolors
\revisioncolorstrue

\begin{document}
 
\title{Collider-observable curvaton}

\author{David Cyncynates\orcidlink{0000-0002-2660-8407}}
\email{davidcyn@ictp.it}
\affiliation{International Centre for Theoretical Physics (ICTP), Strada Costiera 11, 34151 Trieste, Italy}
\affiliation{Istituto Nazionale di Fisica Nucleare (INFN), Sezione di Trieste, I-34127 Trieste, Italy}

\author{Takuya Okawa\orcidlink{0009-0003-9478-380X}}
\email{tokawa@sissa.it}
\affiliation{SISSA, International School for Advanced Studies, via Bonomea 265, 34136, Trieste, Italy}
\affiliation{IFPU - Institute for Fundamental Physics of the Universe, Via Beirut 2, 34014 Trieste, Italy}
\affiliation{Istituto Nazionale di Fisica Nucleare (INFN), Sezione di Trieste, I-34127 Trieste, Italy}

\date{\today}
\preprint{SISSA 21/2026/FISI}

\begin{abstract}
We study a technically natural realization of the curvaton mechanism, in which decay is mediated through the superrenormalizable interaction with the Higgs. We compute the quantum and thermal corrections to the curvaton potential and show that they remain under control throughout its cosmological evolution. For high inflationary scales, the curvaton dissipates through off-shell processes in the Standard Model plasma. The viable parameter space is broad, and the region in which the curvaton mass and dimensionful Higgs-portal coupling are closest in scale lies within the discovery reach of proposed searches for light Higgs-mixed scalars.
\end{abstract}

\maketitle

\section{Introduction}
\label{sec: introduction}

Cold single-field inflation provides an economical account of the observed homogeneity, isotropy, and near scale invariance of the primordial perturbations~\cite{Starobinsky:1980te,Guth:1980zm,Linde:1981mu,Mukhanov:1981xt}. Even so, fitting all observational requirements with a single scalar can be challenging. The inflaton must both drive a sufficiently long period of accelerated expansion and generate the scalar perturbations consistent with the measured amplitude, the red tilt, and the upper bounds on non-Gaussianity~\cite{Planck:2018vyg, Planck:2018jri, Planck:2019kim}. 
This has motivated inflationary model building where the origin of the curvature perturbation is partially or fully separated from the dynamics of the inflaton itself.

A particularly simple realization of this idea is the curvaton mechanism~\cite{Lyth:2001nq,Enqvist:2001zp,Moroi:2001ct,Bartolo:2002vf}, in which the nearly scale-invariant fluctuations of a light spectator field are subsequently converted into adiabatic curvature perturbations. During inflation, the curvaton $\sigma$ has a non-zero expectation value, remains energetically subdominant while its perturbations are generated, and later comes to make up an appreciable fraction of the total energy density before decaying to Standard Model particles. In the minimal treatment, the scenario is largely characterized by four quantities: the curvaton field value $\sigma_\star$ at the horizon exit of the pivot scale $k_\star$, its decay rate $\Gamma_\sigma$, its mass $m_\sigma$, and the inflationary scale $H_\mathrm{inf}$.

This parametrization is useful, but it can obscure important consistency conditions. The curvaton cannot simply be assigned an arbitrary lifetime: its decay is mediated by an interaction with Standard Model particles, and that same interaction necessarily affects its evolution. At zero temperature, it generates radiative corrections to the curvaton potential, modifying its mass and self-interactions~\cite{Coleman:1973jx}. In the primordial plasma, it induces additional thermal corrections, including a tadpole for interactions linear in the curvaton that displaces the field from its zero-temperature minimum, as well as thermal masses and higher-point interactions~\cite{Dolan:1973qd,Weinberg:1974hy,Batell:2021xsi,Batell:2022hsf,Cyncynates:2024bxw,Cyncynates:2024ufu}. Finally, the interaction gives rise to dissipative processes that transfer energy from the curvaton to the plasma, potentially even when perturbative decays are kinematically forbidden~\cite{Bodeker:2022ihg,Banerjee:2025qyg}. A self-consistent realization must therefore account for vacuum corrections, thermal corrections, and damping when determining the curvaton's evolution and eventual decay.

Several particle-physics realizations of the curvaton mechanism have been explored, including supersymmetric flat directions and pseudo-Nambu--Goldstone bosons~\cite{Enqvist:2003mr,Dimopoulos:2003az}. These constructions can also connect the origin of primordial perturbations to other questions in particle physics, including the strong CP problem through Peccei--Quinn dynamics~\cite{Dimopoulos:2003ii,Chun:2004gx}, neutrino masses and leptogenesis through right-handed sneutrinos~\cite{Mazumdar:2003bs}, and the baryon asymmetry through Affleck--Dine dynamics~\cite{Ireland:2024wwn}. Higgs-coupled curvatons have also been studied, particularly through quartic interactions, for which radiative corrections and the thermal blocking of resonant Higgs production can significantly affect the curvaton's evolution and decay~\cite{Enqvist:2012tc,Enqvist:2013qba,Enqvist:2013gwf}. More generally, thermal corrections and dissipation have been shown to restrict the viable curvaton parameter space and affect the conversion of its fluctuations into adiabatic curvature perturbations~\cite{Mukaida:2014yia,Kitajima:2014xna}.

In this work, we study the curvaton scenario with the superrenormalizable interaction between a gauge-singlet real scalar curvaton field $\sigma$ and
the Standard Model Higgs boson $H$,
\begin{align}
    \mathcal L\supset -\mu\sigma H^\dagger H\,.\label{eqn:interaction}
\end{align}
The superrenormalizability keeps the quantum and thermal corrections to the curvaton and Higgs potentials under control~\cite{Piazza:2010ye,Fradette:2018hhl,Batell:2022hsf}, while off-shell damping provides an efficient means to transfer its energy to the Standard Model before the electroweak phase transition. We show that the primordial curvaton displacement is not significantly affected by thermal misalignment, so that its initial condition after inflation corresponds directly to its energy density at decay. We identify regimes in which the principal curvaton dissipation channels are off-shell damping and perturbative decay, and check that thermal corrections to its potential remain small. We then determine the resulting curvaton parameter space and show that it is probed by future measurements of primordial non-Gaussianity and tensor perturbations, and proposed laboratory searches for light Higgs-mixed scalars.\footnote{Throughout this paper, we utilize the following conventions. The Higgs mass parameter in the symmetric phase is $|m_H|\sim 88\GeV$, the electroweak crossover temperature $T_{\rm ew} = 160\GeV$, and in the broken phase it is $m_h = 125.11\GeV$~\cite{ATLAS:2023oaq}. We work in a Friedmann--Lema\^itre--Robertson--Walker (FLRW) cosmology with mostly negative signature. We take the reduced Planck mass $\Mpl = 2.435\times 10^{18}\GeV$ and the Hubble rate at matter radiation equality $H_{\rm eq} = 2.25\times 10^{-28}\eV$~\cite{Planck:2018vyg}.
}

\section{Cosmological evolution}
\label{sec:cosmology}

In this section, we discuss the chronology of the curvaton model,
\begin{align}
    {\cal L} = {\cal L}_{\rm SM} + \frac12(\partial\sigma)^2 - \frac12m_\sigma^2\sigma^2 - \mu\sigma H^\dagger H\,.
\end{align}
In \cref{subsec:initial_conditions}, we outline the requirements on the curvaton initial condition so that it can reproduce the observed CMB temperature anisotropy, and in \cref{subsec:damping} we review the mechanisms by which the curvaton transfers its energy density to the Standard Model bath and check that perturbation theory remains valid over the entirety of its cosmological evolution. Precise expressions are provided in~\cref{app:thermal_loop_resummation,app:damping,app:equilibration,app:decay}.

\subsection{Initial conditions}
\label{subsec:initial_conditions}

While the curvaton is light compared to the scale of inflation $m_\sigma\ll H_{\rm inf}$, it acquires approximately Gaussian fluctuations with power spectrum $\langle\delta\sigma(\mathbf k)\delta\sigma(\mathbf k')\rangle\equiv(2\pi)^3\delta^{(3)}(\mathbf k+\mathbf k')(2\pi^2/k^3)\mathcal P_{\delta\sigma_\star}(k)$,
\begin{align}
    \mathcal P_{\delta\sigma_\star}\simeq\left(\frac{H_{\rm inf}}{2\pi}\right)^2,
\end{align}
uncorrelated with the inflaton field fluctuations. We assume that the observed red tilt is generated by a slow decrease of the Hubble rate during inflation, $n_s-1\simeq 2\dot H_\star/H_\star^2$, with a negligible contribution from the curvaton mass. In this case, the inflaton itself generates a curvature perturbation ${\cal P}\simeq 2.8\times 10^{-10}(H_{\rm inf}/4.7\times 10^{13}\GeV)^2$, consistently smaller than the observed curvature perturbation. Therefore, in our estimates below, we neglect the inflaton-sourced radiation curvature perturbation, so that the curvaton accounts for the observed scalar power.\footnote{For the largest allowed inflationary scales, the inflaton introduces at most an ${\cal O}(10\%)$ correction.}

Within the separate Universe approximation~\cite{Salopek:1990jq,Wands:2000dp,Rigopoulos:2003ak}, the local curvaton density is determined by its primordial field value
\begin{align}\label{eq:curvaton_transfer_function}
    F(\sigma_\star)\equiv \rho_\sigma(t_{\rm dec},\sigma_\star)|_{\rm flat}\,,
\end{align}
where $t_{\rm dec}$ is the time when the curvaton damps/decays, and the density is evaluated on a spatially flat reference slice.

In the sudden-decay approximation, the conversion efficiency is
\begin{align}
    \label{eq:curvaton_Rsigma}
    R_\sigma\equiv\left.\frac{3\bar\rho_\sigma}{4\bar\rho_r+3\bar\rho_\sigma}\right|_{\rm dec}\,,
\end{align} 
and the scalar power and local non-Gaussianity of the resulting radiation perturbations are respectively
\begin{subequations}
\begin{align}\label{eq:curvaton_As}
    A_s&=\left[\frac{R_\sigma H_\star}{6\pi}\frac{F'}{F}\right]^2\,,\\
    \label{eq:curvaton_fNL}
    f_{\rm NL}&=\frac{5}{2R_\sigma}\frac{F F''}{F'^2}-\frac{5}{3}-\frac{5R_\sigma}{6}\,,
\end{align}
\end{subequations}
where $'$ denotes differentiation with respect to $\sigma_\star$.

For the quadratic curvaton potential
\begin{align}\label{eqn:bare_potential}
    V(\sigma)=\frac{1}{2}m_\sigma^2\sigma^2\,,
\end{align}
with a freely chosen initial displacement, we have $F\propto\sigma_\star^2$.  Hence $F'/F=2/\sigma_\star$ and $F F''/F'^2=1/2$, giving~\cite{Lyth:2002my}
\begin{subequations}
\begin{align}\label{eqn:As_usual_curvaton}
    A_s&=\left(\frac{R_\sigma H_\star}{3\pi\sigma_\star}\right)^2\,,\\
    f_{\rm NL}&=\frac{5}{4R_\sigma}-\frac{5}{3}-\frac{5R_\sigma}{6}\,.
\end{align}
\end{subequations}
The observed $A_s\simeq2.1\times10^{-9}$~\cite{Planck:2018jri} then requires
\begin{align}\label{eq:curvaton_sigma_star}
    \sigma_\star\simeq2.3\times10^3 R_\sigma H_\star\,.
\end{align}
The local non-Gaussianity bound $f_{\rm NL}^{\rm local} = -0.9\pm 5.1$~\cite{Planck:2019kim} prevents $R_\sigma$ from being too small. Using the quoted $68\%$ uncertainty below gives $R_\sigma\gtrsim0.21$.

The energy density fraction at decay thus determines both the amplitude of adiabatic perturbations and the non-Gaussianity through $R_\sigma$. Assuming radiation domination at the onset of curvaton oscillations at $H\sim m_\sigma$ and negligible entropy production before decay, the resulting curvaton fraction is, using \cref{eqn:As_usual_curvaton},\footnote{Since all dynamics occur long before neutrino decoupling, we will neglect differences between the relativistic and entropy degrees of freedom.}
\begin{equation}
\begin{aligned}\label{eq:curvaton_Xi_dec_As}
    \left.\frac{\rho_\sigma}{\rho_r}\right|_{\rm dec}\approx& \frac{4\Gamma^2(5/4)}{27\pi^3}\frac{R_\sigma^2}{A_s}\left(\frac{g_\star(T_{\rm dec})}{g_\star(T_{\rm osc})}\right)^{1/4}\\&\times\left(\frac{H_\star}{\Mpl}\right)^2\sqrt{\frac{m_\sigma}{H_{\rm rad}(T_{\rm dec})}}\,,
\end{aligned}
\end{equation}
where $H_{\rm rad}(T) \equiv \sqrt{\pi^2g_\star/90}T^2/\Mpl$ and $T_{\rm dec}$ are the radiation-only Hubble rate and temperature at the time of curvaton decay.
Provided the curvaton is sufficiently (co-)dominant, the resulting non-Gaussianity is ${\cal O}(1)$, within the observationally allowed range~\cite{Planck:2019kim}.

\subsection{Damping and decay}
\label{subsec:damping}

The oscillating curvaton field transfers energy to the Standard Model through two main channels: off-shell damping and on-shell decay. Off-shell damping occurs because curvaton oscillations modulate the Higgs mass, causing its thermal distribution to relax toward a changing equilibrium and dissipate energy in the process~\cite{Bodeker:2022ihg,Banerjee:2025qyg}. Unlike on-shell decay, this mechanism does not require kinematically accessible decay channels. Further, perturbative decay is thermally blocked when the in-medium masses of the final-state particles exceed the curvaton mass, meaning off-shell damping is often the only accessible dissipation channel.

Both processes may be incorporated into a single effective damping rate
\begin{align}
    \Upsilon_{\rm tot} = \Upsilon_{\rm off} + \Upsilon_{\rm on}\,.
\end{align}
For underdamped curvaton oscillations, the integrated damping determines the characteristic time $t_{\rm dec}$ at which its energy is dissipated:
\begin{align}\label{eqn:decay_time}
    \int_{t_{\rm osc}}^{t_{\rm dec}}{\rm d}t'\,\Upsilon_{\rm tot}\approx 1\,,
\end{align}
where $t_{\rm osc}\sim 1/m_\sigma$ is the onset of curvaton oscillations. Since the early-time contribution is negligible, the lower limit is extended to zero in our estimates.

The off-shell damping rate scales parametrically as~\cite{Banerjee:2025qyg}
\begin{align}\label{eqn:Upsilon_off}
    \Upsilon_{\rm off}\sim\frac{\mu^2}{8\pi}\frac{\Gamma_{\rm th}}{\Gamma_{\rm th}^2+m_\sigma^2}
    \begin{cases}
        \dfrac{1}{g_{\rm eff}}, & T\gg T_{\rm ew},\\\\
        \dfrac{T^4}{m_h^4}, & T\ll T_{\rm ew},
    \end{cases}
\end{align}
where $T_{\rm ew}$ denotes the electroweak crossover temperature and $m_h$ the Higgs vacuum mass, $g_{\rm eff}^2\approx0.4$ parametrizes the Higgs thermal mass $m_H(T)\approx g_{\rm eff}T$, and $\Gamma_{\rm th}$ is the Higgs equilibration rate $\Gamma_{\rm th}\approx(0.01{-}0.05)T$ (see \cref{app:thermal_loop_resummation,app:equilibration}). Below electroweak symmetry breaking, Higgs mixing allows the curvaton to modulate the masses and interactions of other Standard Model particles. Estimating the dominant response as arising from thermally populated species with masses of order $T$ gives the additional suppression $(T/m_h)^4$, with $\Gamma_{\rm th}\sim{\cal O}(T)$. This broken-phase expression is a schematic estimate, with the detailed rates and approximations discussed in \cref{app:damping}. 

In general, a temperature-dependent decay rate has been found to change the transfer of perturbations from the curvaton to the thermal bath relative to the sudden decay approximation~\cite{Kitajima:2014xna}. 
However, the symmetric-phase off-shell damping rate grows as temperature decreases while $\Gamma_\mathrm{th} > m_\sigma$, meaning the off-shell damping of the curvaton energy density can be more sudden than a constant on-shell decay rate. We therefore expect the sudden decay approximation to hold quite well, though a detailed check is left to future work.

The on-shell decay rate scales parametrically as
\begin{align}
    \Upsilon_{\rm on}\sim \frac{\mu^2}{8\pi m_\sigma}\Theta\left[\frac{m_\sigma}{T} - 1\right]
    \begin{cases}
        1, & \dfrac{T}{T_{\rm ew}}{\,\rm or\,}\dfrac{m_\sigma}{m_h}\gg 1,\\\\
        \dfrac{m_\sigma^4}{m_h^4}, & \dfrac{T}{T_{\rm ew}}{\,\rm and\,}\dfrac{m_\sigma}{m_h}\ll 1,
    \end{cases}
\end{align}
where the step function schematically accounts for thermal blocking. The first expression describes direct decay into Higgs excitations, while the second gives the characteristic scaling of decay through Higgs mixing. The detailed decay rates and thermal thresholds are discussed in \cref{app:decay}.

Two limiting behaviors help illustrate the resulting cosmological evolution. For sufficiently large portal couplings, the curvaton dissipates before electroweak symmetry breaking, when Higgs excitations are abundant. In the adiabatic regime, $\Gamma_{\rm th}\gg m_\sigma$, the off-shell damping rate scales as $\Upsilon_{\rm off}\propto \mu^2/T$, so that $\Upsilon_{\rm off}/H\propto T^{-3}$ during radiation domination. Dissipation therefore becomes efficient rapidly as the Universe cools, giving
\begin{align}\label{eqn:T_dec_off}
    T_{\rm dec}\sim\left(\mu^2\Mpl\right)^{1/3}\sim \TeV\left(\frac{\mu}{100\keV}\right)^{2/3}\,.
\end{align}
For sufficiently small portal couplings, the curvaton instead survives electroweak symmetry breaking and decays much later through Higgs mixing, when thermal blocking has become negligible. In this limit,
\begin{align}\label{eqn:T_dec_on}
    T_{\rm dec}\sim\left(\frac{\mu^2m_\sigma^3\Mpl}{m_h^4}\right)^{1/2}\sim 10\MeV\left(\frac{\mu}{\rm keV}\right)\left(\frac{m_\sigma}{{\rm GeV}}\right)^{3/2}\,.
\end{align}
Intermediate couplings interpolate between these limits.

Because the curvaton initial misalignment is determined by the scale of inflation and the CMB temperature perturbations, requiring $R_\sigma={\cal O}(1)$ at decay fixes the scale of inflation. Larger couplings imply earlier decay, leaving less time for the curvaton energy fraction to grow relative to radiation. Therefore, larger coupling requires a larger initial displacement and, correspondingly, a higher inflationary scale. In the limits of \cref{eqn:T_dec_off,eqn:T_dec_on}, and assuming instantaneous reheating, we obtain respectively (fixing $R_\sigma = 0.5$)
\begin{subequations}\label{eqn:H_star}
\begin{align}
    H_\star&\sim 6\times 10^{12}\GeV\left(\frac{\mu}{100\keV}\right)^{1/3}\left(\frac{m_\sigma}{\rm GeV}\right)^{-1/4}\,,\\
    H_\star&\sim 2\times 10^{10}\GeV\left(\frac{\mu}{\rm keV}\right)^{1/2}\left(\frac{m_\sigma}{\rm GeV}\right)^{1/2}\,.
\end{align}
\end{subequations}

The initial curvaton displacement can therefore approach the GUT scale, making it necessary to verify that its effect on the Higgs remains perturbative throughout the cosmological evolution. At temperatures well above electroweak symmetry breaking, this requires that the curvaton-induced shift in the Higgs mass parameter be small compared with its thermal contribution,
\begin{align}\label{eqn:perturbativity}
    |\mu\sigma|\ll g_{\rm eff}^2T^2\,.
\end{align}
For a quadratically oscillating curvaton, the energy density is $\rho_\sigma=m_\sigma^2\bar\sigma^2/2$, where $\bar\sigma$ denotes the oscillation envelope. Comparing with the radiation density $\rho_r=\pi^2g_\star T^4/30$, we obtain
\begin{align}\label{eqn:perturbativity_check}
    \epsilon_H(T)
    \equiv\frac{|\mu|\bar\sigma}{g_{\rm eff}^2T^2}
    \approx20\sqrt{\frac{\rho_\sigma}{\rho_r}}\frac{|\mu|}{m_\sigma}\,.
\end{align}
Deep in the broken phase, the appropriate comparison is instead with the vacuum Higgs mass squared, $m_h^2\simeq(125\GeV)^2$. Through the electroweak crossover, the relevant comparison is with the restoring curvature of the Higgs effective potential about its thermal equilibrium configuration. Using the screening mass in Fig.~8 of Ref.~\cite{DOnofrio:2015gop} as an estimate of this curvature gives a condition on $|\mu\sigma|$ that is $\mathcal O(10)$ more restrictive than either the zero-temperature or high-temperature estimates.

\subsection{Thermal misalignment}
\label{subsec:thermal_misalignment}
A final consistency check is whether thermal misalignment~\cite{Batell:2021xsi,Batell:2022hsf,Cyncynates:2024bxw,Cyncynates:2024ufu} plays a role in the curvaton dynamics. As we discuss in \cref{app:thermal_loop_resummation}, so long as $\sigma$ satisfies the perturbativity condition \cref{eqn:perturbativity}, the leading correction to its effective potential at high temperature is a linear term
\begin{align}
    \Delta V_{\rm eff}\sim \frac{\mu T^2}{6}\sigma\,.
\end{align}
This induces an effective initial displacement 
\begin{align}
    \sigma_{\rm th} \sim -\frac{0.2}{\sqrt{g_\star}}\frac{\Mpl\mu}{m_\sigma}\,,
\end{align}
which we then compare to the initial misalignment required to produce the observed curvature perturbation. If decay occurs in the symmetric phase or broken phase limits of \cref{eqn:H_star} respectively, we find
\begin{subequations}
\begin{align}
    \frac{\sigma_{\rm th}}{\sigma_\star}&\sim 6\times 10^{-4}\left(\frac{\mu}{100\keV}\right)^{2/3}\left(\frac{m_\sigma}{\rm GeV}\right)^{-3/4}\,,\\
    \frac{\sigma_{\rm th}}{\sigma_\star}&\sim 2\times 10^{-3}\left(\frac{\mu}{\rm keV}\right)^{1/2}\left(\frac{m_\sigma}{\rm GeV}\right)^{-3/2}\,,
\end{align} 
\end{subequations}
indicating that thermal misalignment is negligible.

These estimates illustrate that the curvaton mechanism can be realized with a superrenormalizable Higgs coupling while maintaining control over the thermal corrections to its evolution. Off-shell damping can efficiently transfer the curvaton's energy to the Standard Model before electroweak symmetry breaking, even for portal couplings much smaller than the curvaton mass. At yet smaller couplings, the curvaton can instead survive to decay perturbatively at later times. We now discuss the resulting parameter space and prospects for experimental searches.

\section{Phenomenology}
\begin{figure*}[t]
    \centering
    \includegraphics[width=\linewidth]{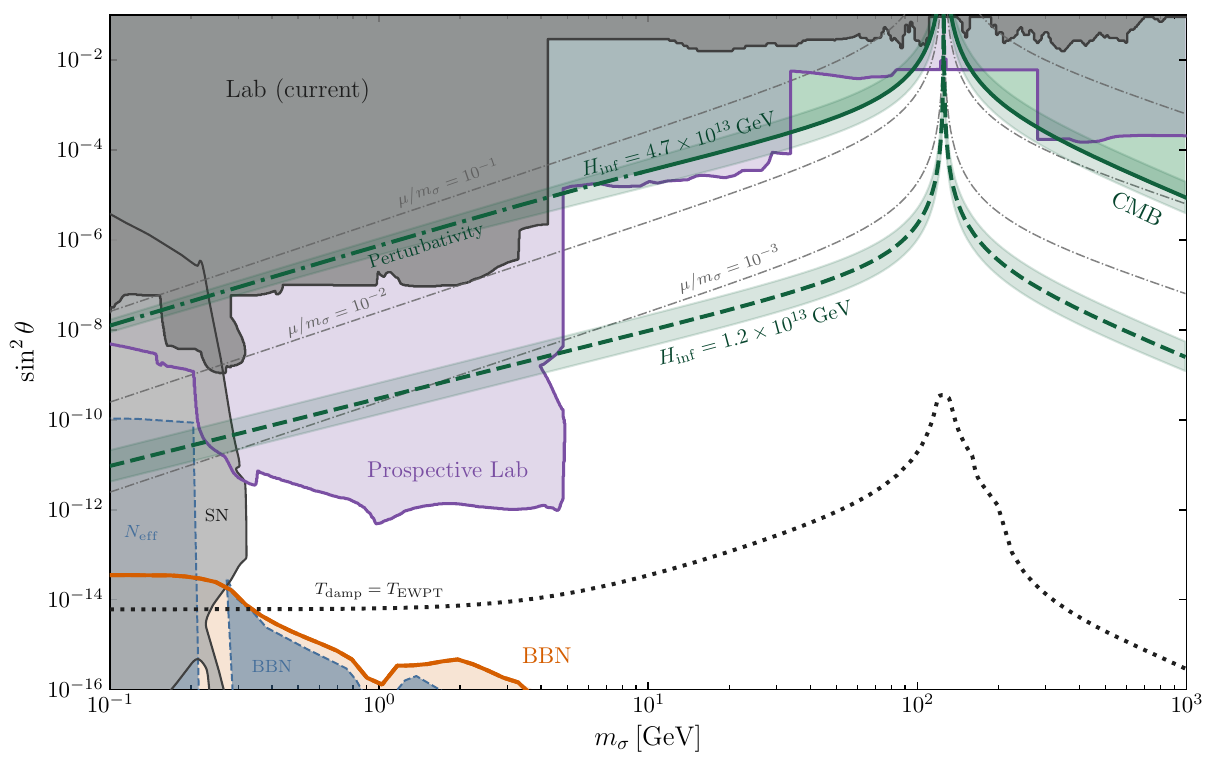}
    \caption{The parameter space of the superrenormalizable Higgs-portal curvaton. The dark-gray shaded region labeled ``Lab'' shows current laboratory exclusions from rare meson decays, proton beam-dump experiments, and neutrino beam experiments~\cite{CHARM:1985anb,Winkler:2018qyg,NA62:2020pwi,NA62:2020xlg,NA62:2021zjw,NA62:2025upx,NA62:2025yzs,MicroBooNE:2022ctm,MicroBooNE:2025jkl,ICARUS:2024oqb,Gorbunov:2021ccu,Foroughi-Abari:2020gju,LHCb:2015nkv,LHCb:2016awg,Belle-II:2023ueh,BaBar:2015jvu} at low masses, Higgs-property measurements by ATLAS~\cite{ATLAS:2022vkf} and CMS~\cite{CMS:2022dwd} at intermediate masses, and collider constraints obtained using \texttt{HiggsTools}~\cite{Bahl:2022igd} at higher masses. The light-gray region labeled ``SN'' shows constraints from SN1987A~\cite{Balaji:2022noj,Hardy:2024gwy,Joseph:2026nut,Yu:2026xtb}. The blue-gray regions show cosmological constraints on the thermal curvaton population from the effective number of relativistic species, ``$N_{\rm eff}$,'' and primordial light-element abundances (``BBN''), assuming a high reheating temperature~\cite{Fradette:2017sdd,Fradette:2018hhl}. The purple shading shows the projected sensitivities of future laboratory searches~\cite{Craik:2022riw,LHCB-FIGURE-2025-001,Jerhot:2936260,Kling:2020mch,Ahdida:2023okr,Alekhin:2015byh,Feng:2022inv,CODEX-b:2019jve,Aielli:2022awh,CODEX-b:2025rck,HIKE:2023ext,Berryman:2019dme,Curtin:2018mvb,Evans:2020aqs,Wang:2019mzd,Canepa:2022swf,Frugiuele:2018coc,EuropeanStrategyforParticlePhysicsPreparatoryGroup:2019qin,deBlas:2025gyz} (see also Refs.~\cite{Alimena:2025kjv,Batell:2022dpx}). The green curves show the upper boundaries consistent with the observed scalar power-spectrum amplitude and the bound on $f_{\rm NL}$~\cite{Planck:2019kim} for $H_{\rm inf}=4.7\times10^{13}\,{\rm GeV}$ (upper) and $H_{\rm inf}=1.2\times10^{13}\,{\rm GeV}$ (lower, dashed). The green shaded region that extends upwards is excluded by the non-observation of tensor perturbations in the CMB. The dot-dashed segment of the upper curve is additionally limited by the perturbativity criterion $\epsilon_H\leq1$; this criterion does not further restrict the lower curve. The shaded green bands show the variation obtained for $\Gamma_{\rm th}/T=0.01$--$0.05$. The orange curve labeled ``BBN'' gives the lower boundary obtained by requiring the temperature of the Universe after energy transfer from the curvaton to the SM to exceed $4\MeV$. The black dotted line corresponds to $T_{\rm damp}=T_{\rm ew}$, and the gray dot-dashed lines indicate constant ratios $\mu/m_\sigma=10^{-3},10^{-2},10^{-1}$.}
    \label{fig:results}
\end{figure*}

The mixing versus mass parameter space of the superrenormalizable Higgs-portal curvaton is shown in \cref{fig:results}. The mixing angle $\theta$ is determined by the underlying portal coupling $\mu$ and the Higgs VEV $v_{\rm ew}\approx246\GeV$,
\begin{align}
    \theta\sim\frac{\mu v_{\rm ew}}{m_\sigma^2-m_h^2}\,.
\end{align}
At large coupling, the curvaton can in principle destabilize the Higgs VEV if $\mu^2\geq 2\lambda_H m_\sigma^2$ and generate a large quantum correction to its mass if $|\mu|/(4\pi)\gtrsim m_h$. In practice, avoiding these limits does not introduce additional constraints on the parameter space in \cref{fig:results} because $\mu$ must be less than ${\cal O}(0.1)m_\sigma$ to satisfy constraints on non-Gaussianity and the scale of inflation. For $m_\sigma\gg {\rm TeV}$, corrections to the Higgs mass become non-trivial, though in this case $m_\sigma$ is beyond the reach of laboratory probes. For reference, we add the gray dot-dashed lines which indicate constant ratios $\mu/m_\sigma=10^{-3},10^{-2},10^{-1}$. 

\paragraph*{Cosmology \& astrophysics.}The upper green contour indicates the curvaton parameter space corresponding to inflationary scales probed by the BICEP/Keck-BAO bound $r\lesssim0.036$~\cite{BICEPKeck:2024stm}, while the lower green contour illustrates the projected sensitivity of the Simons Observatory, $r\sim0.0023$~\cite{SimonsObservatory:2025avm}. At the largest inflationary scales and smaller $m_\sigma$, maintaining perturbativity mildly strengthens the inflationary bound, indicated by dot-dashing. See \cref{app:curvaton_cosmological_bounds} for further details.

At fixed inflationary scale, the parameter space is bounded above by the requirement that the curvaton mechanism realizes the observed amplitude of the scalar power spectrum while not having excessive non-Gaussianity. Larger couplings lead to earlier dissipation of the curvaton energy density, leaving less time for the curvaton energy fraction to grow relative to radiation. Cosmological probes will measure primordial non-Gaussianity, probing the curvaton energy fraction at decay~\cite{Karagiannis:2019jjx,Heinrich:2023qaa,Euclid:2024ris}.
At sufficiently small couplings, the curvaton survives until late times, at which point its decay during or after Big Bang Nucleosynthesis (BBN) would spoil the predicted abundances of the light elements: the orange contour illustrates this lower bound on the coupling, obtained by requiring the temperature of the SM bath after energy transfer from the curvaton sector to exceed $T_\mathrm{BBN} \simeq 4 \MeV$. Separately, the thermal production of the curvaton from the SM bath is constrained by its effect on BBN and from its contribution to $\Delta N_{\rm eff}$~\cite{Fradette:2017sdd,Fradette:2018hhl}, which we indicate by the gray dashed regions labeled ``BBN'' and ``$N_{\rm eff}$'' in \cref{fig:results}. \Cref{app:curvaton_cosmological_bounds} further details the assumptions that determine these boundaries.

A stronger lower bound on the coupling can arise if the baryon asymmetry is generated before the curvaton transfers its perturbations to radiation. In this case, baryons and photons generally inherit different curvature perturbations, producing correlated baryon isocurvature strongly constrained by the CMB~\cite{Planck:2018jri}. Although weaker than CMB bounds for scale-invariant isocurvature, BBN constraints on baryon isocurvature extend to much smaller comoving scales through the effect of baryon inhomogeneities on primordial light-element abundances~\cite{Inomata:2018htm,Bagherian:2025puf}. Assuming baryogenesis occurs independently of the curvaton at or above the electroweak scale, the parameter space below the black-dotted line is disfavored. Earlier baryogenesis requires correspondingly earlier curvaton decay and therefore larger couplings. For illustration, requiring $T_{\rm dec}\gtrsim10^8\GeV$, as motivated by high-scale leptogenesis~\cite{Buchmuller:2004tu,Davidson:2002qv,Giudice:2003jh}, gives $\mu\gtrsim3\TeV$ from \cref{eqn:T_dec_off}, pushing the viable curvaton masses well beyond the range of conventional laboratory searches, and posing a naturalness problem for the Higgs.

Observations of SN1987A constrain light scalars mixing with the Standard Model. Scalars produced in the hot proto-neutron-star core can carry away energy if they escape without being reabsorbed or decaying, thereby shortening the neutrino signal observed from the supernova. Requiring that this additional cooling by the curvaton remain consistent with observations excludes part of the low-mass parameter space labeled ``SN''~\cite{Balaji:2022noj,Hardy:2024gwy,Joseph:2026nut,Yu:2026xtb}.

\paragraph*{Laboratory.}Existing laboratory searches constrain the upper gray region of \cref{fig:results}. Scalars mixing with the Higgs can be produced through rare meson decays, fixed-target and beam-dump experiments, and high-energy collisions with subsequent decays to visible Standard Model particles~\cite{Winkler:2018qyg,Ferber:2023iso}. At low masses, rare kaon decays provide robust bounds through searches at NA62~\cite{NA62:2020pwi,NA62:2020xlg,NA62:2021zjw,NA62:2025upx}, and rare $B$-meson decays at LHCb~\cite{LHCb:2015nkv,LHCb:2016awg}, Belle II~\cite{Belle-II:2023ueh}, and BaBar~\cite{BaBar:2015jvu}. Complementary bounds are obtained by proton beam-dump experiments such as CHARM~\cite{CHARM:1985anb,Winkler:2018qyg}, PS191~\cite{Gorbunov:2021ccu}, and NA62 in beam-dump mode~\cite{NA62:2025yzs}, as well as neutrino beam experiments including LSND~\cite{Foroughi-Abari:2020gju}, MicroBooNE~\cite{MicroBooNE:2022ctm,MicroBooNE:2025jkl}, and ICARUS~\cite{ICARUS:2024oqb}. 

At intermediate masses, measurements of Higgs properties by ATLAS~\cite{ATLAS:2022vkf} and CMS~\cite{CMS:2022dwd} mainly place bounds on the mixing parameter. The mixing between the scalar and the Higgs suppresses the couplings of the observed Higgs boson to SM particles universally by a factor of $\cos\theta$, leading to a suppression of the signal strength by $\cos^2\theta$. In our setup, the Higgs can also decay into a pair of scalars when $m_\sigma < m_h/2$. Thus, we reinterpret the bounds on the Higgs signal strength, defined as the ratio of the observed Higgs production rate (multiplied by a branching ratio) to the SM prediction, by including this decay channel:
\begin{align}
    \mu_h \equiv \frac{\cos^4\theta~\Gamma_h^\mathrm{SM}}{\cos^2\theta~\Gamma_h^\mathrm{SM} + \Gamma(h \to \sigma\sigma)}\,.
\end{align}
At higher masses, direct searches for an additional Higgs-like scalar place stringent bounds, which we evaluate by using {\tt HiggsTools}~\cite{Bahl:2022igd}.

Future laboratory experiments will probe smaller Higgs mixing angles and potentially intersect the parameter space consistent with the curvaton cosmology. The prospective sensitivity is illustrated in the purple regions of \cref{fig:results}. At low masses, future searches include LHCb~\cite{Craik:2022riw,LHCB-FIGURE-2025-001}, NA62~\cite{Jerhot:2936260}, FASER~\cite{Kling:2020mch}, SHiP~\cite{Ahdida:2023okr,Alekhin:2015byh}, FASER2~\cite{Feng:2022inv}, CODEX-b~\cite{CODEX-b:2019jve,Aielli:2022awh,CODEX-b:2025rck}, HIKE~\cite{HIKE:2023ext}, DUNE~\cite{Berryman:2019dme}, MATHUSLA~\cite{Curtin:2018mvb}, and CMS~\cite{Evans:2020aqs}. These experiments primarily probe light scalars produced in rare meson decays through the displaced vertices of their subsequent visible decays. At intermediate masses, ILC, CEPC, FCC-ee, and Tera-Z~\cite{Wang:2019mzd,Canepa:2022swf,Frugiuele:2018coc,EuropeanStrategyforParticlePhysicsPreparatoryGroup:2019qin} constrain the mixing through measurements of Higgs properties and direct searches for scalars. At high masses, additional sensitivity can be achieved by HL-LHC, FCC-ee, FCC-hh, LCF-250, CLIC, LEP3, and MuC~\cite{deBlas:2025gyz}. 

\section{Discussion}
\label{sec:discussion}

We have studied a realization of the curvaton mechanism in which a light scalar couples to the Standard Model through a superrenormalizable Higgs interaction. The interaction allows the curvaton to transfer its energy to the Standard Model through off-shell thermal damping or perturbative decay, while the quantum and thermal corrections to its potential remain under control. We find a broad parameter space compatible with the observed curvature perturbations, including regions within reach of future searches for Higgs-mixed scalars.

An important motivation for this setup is its technical naturalness, as previously emphasized in studies of the superrenormalizable Higgs portal~\cite{Piazza:2010ye,Fradette:2018hhl,Batell:2022hsf}. The curvaton possesses an approximate shift symmetry, softly broken by its mass and the Higgs coupling. The interaction generates radiative mass corrections $\delta m_\sigma^2,\delta m_h^2\sim\mu^2/(16\pi^2)$, up to logarithms, so that requiring $\mu^2/(16\pi^2)\ll m_\sigma^2,m_h^2$ ensures that neither the curvaton nor Higgs receives a large radiative mass correction. The quadratic divergence instead appears in a curvaton tadpole proportional to $\langle H^\dagger H\rangle$. This tadpole can be absorbed by a field redefinition, which shifts the Higgs mass parameter; its implications for naturalness therefore depend on the ultraviolet completion of the electroweak sector.

The curvaton mechanism has motivated a variety of particle-physics realizations, including axion-like particles, supersymmetric flat directions, and Higgs-coupled scalars~\cite{Enqvist:2003mr,Dimopoulos:2003az,Chun:2004gx,Enqvist:2013gwf}. Axion-like curvatons enjoy similar symmetry protection, but the large primordial field displacement required to generate the observed curvature perturbations typically implies a large axion decay constant, suppressing their interactions with Standard Model particles. In our model, the initial displacement and coupling are independent parameters, allowing the curvaton to retain experimentally accessible interactions. Previous studies have also considered curvatons with quartic Higgs interactions~\cite{Enqvist:2012tc,Enqvist:2013qba,Enqvist:2013gwf}. The superrenormalizable coupling considered here provides a simple realization in which the same interaction determines both cosmological dissipation and laboratory phenomenology.

Future laboratory searches for light Higgs-mixed scalars may therefore provide a direct test of the curvaton mechanism, complementary to cosmological measurements of primordial non-Gaussianity and tensor perturbations. The viable parameter space also depends on the cosmological history, particularly the origin of the baryon asymmetry, which can impose additional consistency conditions through residual isocurvature perturbations. A Higgs-mixed scalar discovery would map directly onto the allowed ranges of cosmological observables and vice versa, offering the possibility that we may discover the particle responsible for the primordial curvature perturbations.

\section*{Acknowledgements}
We are grateful to Mehrdad Mirbabayi for insightful comments on the draft. We also thank Rouzbeh Allahverdi, Aleksandr Azatov, Kim Berghaus, Takeshi Kobayashi, Gordan Krnjaic, Ngan Nguyen, Toby Opferkuch, Rudin Petrossian-Byrne, Nicklas Ramberg, Olivier Simon, Jessie Shelton, Kuver Sinha, Stefan Stelzl, Erwin Tanin, Giovanni Villadoro, Zachary Weiner, and Huangyu Xiao for useful discussions. D.C.\ acknowledges the receipt of a grant from the Abdus Salam International Centre for Theoretical Physics (ICTP), Trieste, Italy, and the support of the Istituto Nazionale di Fisica Nucleare (INFN). T.O.\ is partly supported by INFN initiatives Theoretical Astroparticle Physics (TAsP). This work was performed in part at the Aspen Center for Physics, which is supported by National Science Foundation grant PHY-2210452. 
For facilitating portions of this research, D.C.\ wishes to acknowledge the Center for Theoretical Underground Physics and Related Areas (CETUP*), The Institute for Underground Science at Sanford Underground Research Facility (SURF), and the South Dakota Science and Technology Authority for hospitality and financial support, as well as for providing a stimulating environment. T.O.\ acknowledges the Fermilab Theory Division for its hospitality and support through the Summer Visitors’ Program. T.O.\ also thanks Washington University in St. Louis, the University of Oklahoma, Pennsylvania State University, and Boston University for their hospitality and support during visits where part of this work was carried out.

Generative AI tools, including ChatGPT versions 5 and 6, were used for language editing, code assistance, and to confirm mathematical results. The authors take full responsibility for the content of the manuscript.

\appendix
\crefalias{section}{appendix}

\section{Thermal corrections and initial conditions}
\label{app:thermal_loop_resummation}

\subsection{Effective potential}
Here $m_H^2(\sigma)=m_{H,0}^2+\mu\sigma$ denotes the Higgs mass parameter in the unresummed potential, while $\Pi_H(T)$ describes its thermal correction. The effective potential of the curvaton in-medium is~\cite{Arnold:1992rz,Quiros:1992ez,Curtin:2016urg}
\begin{equation}
\begin{aligned}\label{eqn:effective_potential}
    V_{\rm eff}(\sigma,T) &= V_0[\sigma] + V_{\rm CW}[m_H^2(\sigma)]+ V_{\rm th}[m_H^2(\sigma),T]\\&\qquad+ V_{\rm ring}[m_H^2(\sigma),\Pi_H,T]\,,
\end{aligned}
\end{equation}
where $V_{\rm CW}$ is the one-loop Coleman-Weinberg term~\cite{Coleman:1973jx},
\begin{align}
    V_{\rm CW}[m_H^2] = \frac{g_H}{2} \int_0^{\infty} \frac{d^4p_{E}}{(2\pi)^4}\log\left[p_E^2 + m_H^2\right]\,.
\end{align}
$V_{\rm th}$ is the one-loop thermal effective potential~\cite{Dolan:1973qd},
\begin{align}
    V_{\rm th}[m_H^2,T] = g_H T\int_0^\infty \frac{d^3p}{(2\pi)^3}\log\left[1 - e^{-T^{-1}(p^2 + m_H^2)^{1/2}}\right]\,,
\end{align}
with $g_H = 4$, the number of degrees of freedom of the Higgs.

The ring diagram resums the IR-divergent contributions to the Matsubara zero mode propagator~\cite{Arnold:1992rz,Curtin:2016urg}
\begin{align}
    V_{\rm ring}[m_H^2,\Pi_H,T] = -\frac{g_HT}{12\pi}\left[(m_H^2 + \Pi_H)^{3/2} - (m_H^2)^{3/2}\right]\,,
\end{align}
with the one-loop high-temperature correction to the Higgs dispersion relation~\cite{Quiros:1999jp}
\begin{align}\label{eqn:in-medium-Higgs-mass}
    \Pi_H(T) = \left(\frac{1}{16}g_1^2 + \frac{3}{16}g_2^2 + \frac{1}{2}\lambda_H + \frac{1}{4}y_t^2\right)T^2\equiv g_{\rm eff}^2 T^2\,.
\end{align}
After renormalizing the vacuum-energy, tadpole, and mass terms, the Coleman-Weinberg potential in the $\overline{\rm MS}$ scheme is
\begin{align}\label{eqn:VCW}
    V_{\rm CW} = \frac{4}{64\pi^2}m_H^4(\sigma)\left[\log\left(\frac{m_H^2(\sigma)}{\bar\mu^2}\right) - \frac32\right]\,,
\end{align}
and the leading high-temperature limit of the thermal potential
\begin{equation}
\begin{aligned}\label{eqn:Vth}
    V_{\rm th} &= \frac{4 }{2 \pi ^2} T^4\left(-\frac{\pi ^4}{45}+\frac{\pi ^2 m_H^2(\sigma)}{12 T^2}-\frac{\pi   [m_H^2(\sigma)]^{3/2}}{6 T^3}\right.\\&\left.\quad-\frac{m_H^4(\sigma) }{32 T^4}\left[\log \left(\frac{m_H^2(\sigma)}{16 \pi ^2 T^2}\right) -{\frac{3}{2}+2 \gamma_E }\right]\right)\,.
\end{aligned}
\end{equation} 
Combining these terms gives the following Higgs-induced high-temperature contribution to \cref{eqn:effective_potential},
\begin{equation}
\begin{aligned}\label{eqn:veff}
\Delta V_{\rm eff}
&=
-\frac{2\pi^2}{45}T^4+\frac{m_H^2(\sigma)}{6}T^2-\frac{T}{3\pi}
\left(m_H^2(\sigma) + \Pi_H\right)^{3/2}
\\&\quad
+\frac{m_H^4(\sigma)}{8\pi^2}
\left[
\ln\!\left(\frac{4\pi T}{\bar\mu}\right)-\gamma_E
\right]\,.
\end{aligned}
\end{equation}
Taking $m_H^2(\sigma)/\Pi_H$ small, we find the following leading term containing non-trivial mass dependence
\begin{align}
    \Delta V_{\rm eff} \to \frac{m_H^2(\sigma) T^2}{6}\left(1 - \frac{3}{\pi}g_{\rm eff}\right)\,,\quad |m_H^2(\sigma)|\ll \Pi_H\,.
\end{align}
Near the electroweak phase transition, $g_{\rm eff}\sim 0.55$, though at higher temperatures it becomes substantially smaller because of the running of the Standard Model parameters.

We also note that the leading correction to the curvaton mass can be extracted from \cref{eqn:veff}, which at high temperature is on the order of $\mu^2/(4\pi g_{\rm eff})$ up to logarithms of the temperature. For the relatively small $\mu/m_\sigma$ values in the allowed parameter space of \cref{fig:results}, this correction is small enough to be neglected.

\subsection{Thermal tadpole and field evolution}
\label{app:homogeneous}

To isolate the conservative thermal forcing, we first consider the quadratic equation without plasma damping,
\begin{align}\label{eqn:EOM}
    \ddot\sigma+3H\dot\sigma+m_\sigma^2\sigma=-\frac16\mu T^2\,.
\end{align}
By linearity, the solution is the sum of the primordial and thermally driven components,
\begin{align}
    \sigma(t)=\sigma_\star u(t)+\sigma_{\rm th}^{(0)}(t)\,,
    \label{eq:primordial_thermal_decomposition}
\end{align}
where $u$ solves the homogeneous equation with $u=1$, $\dot u=0$ initially, and $\sigma_{\rm th}^{(0)}$ has vanishing initial displacement and velocity.
\cref{eqn:EOM} admits an analytic Green function solution. While $H\gg m_\sigma$, the thermal force displaces the curvaton,
\begin{align}
    \sigma = 
    -\frac{5}{2\pi^2 g_\star(T_{\rm osc})}
    \mu\frac{\Mpl^2}{T^2}
    +\sigma_{\star}\,.
\end{align}
Oscillations begin once $3H\simeq m_\sigma$. For the thermally driven solution alone, the maximum displacement is
\begin{align}\label{eqn:sigma_max}
    \sigma_{\rm th,max}^{(0)}
    =
    -\frac{0.226}{\sqrt{g_\star(T_{\rm osc})}}
    \frac{\Mpl\mu}{m_\sigma}\,.
\end{align} 
The thermal solution quantifies the correction to the primordial curvaton oscillations. Its late-time oscillation energy density is
\begin{align}\label{eqn:relic_abundance}
    \rho_{\sigma,\rm th}^{(0)}=\frac{5\,\Gamma^2(\frac34)}{24\sqrt3\,\pi^2\,g_\star(T_{\rm osc})}\mu^2M_{\rm pl}^2\left(\frac{a_{\rm osc}}{a}\right)^3\,,
\end{align}
where $a_{\rm osc}$ denotes the scale factor at $3H=m_\sigma$.

It is worth noting that the thermal misalignment piece and the initial misalignment piece are exactly 90 degrees out of phase with one another, and hence their energy densities add. This is also a reason why thermal misalignment on its own cannot supply the necessary curvature perturbation without (self)-interactions to misalign the phases~\cite{Batell:2026sml}.

\section{Thermal damping}
\label{app:damping}

As the homogeneous curvaton field oscillates with a frequency $\omega \approx m_\sigma$, it causes the mass of the Higgs to change adiabatically. In response, the Higgs thermal distribution function 
\begin{align}
    f_p^{\rm eq}(T) \equiv \frac{1}{e^{\sqrt{m_H^2 + p^2}/T} - 1}\,,
\end{align} 
must readjust to maintain local thermal equilibrium. In relaxing to its new thermal equilibrium, it dissipates energy into the plasma via number changing processes. Equivalently, one may calculate the damping rate directly from the scattering of homogeneous $\sigma$ as in Ref.~\cite{Bodeker:2022ihg}.

The formalism of Ref.~\cite{Banerjee:2025qyg} gives an elegant path to calculate this damping rate. From their Eq.~(23), the adiabatic damping rate is
\begin{align}
    \Upsilon^{\rm ad} &= \frac{2[m_H\partial_\sigma m_H]^2}{\pi^2\Gamma_{\rm th}T}\int\frac{p^2{\rm d} p}{m_H^2 + p^2}f_p^{\rm eq}\left[1 + f_p^{\rm eq}\right]\,,
\end{align}
where $\Gamma_{\rm th}$ is the Higgs thermal equilibration rate, which we estimate in \cref{app:equilibration} to be
\begin{align}
    \Gamma_{\rm th}\sim (0.01{-}0.05) T\,,
\end{align}
and we have used the thermally corrected Higgs dispersion relation
\begin{align}
    m_H^2 = m_{H,0}^2 + \Pi_H(T),
\end{align}
with $m_{H,0}^2 \approx -(88\GeV)^2$ and the thermal self-energy $\Pi_H(T)$ given in \cref{eqn:in-medium-Higgs-mass}, where $g_{\rm eff}^2\approx0.4$.
In the high-temperature limit where both $m_\sigma$ and $m_{H,0}$ can be neglected relative to $T$, we obtain
\begin{align}
    \Upsilon^{\rm ad} = \frac{f(g_{\rm eff})}{8\pi g_{\rm eff}}\frac{\mu^2}{\Gamma_{\rm th}}\,,
\end{align}
where
\begin{align}
    f(g_{\rm eff}) = \frac{g_{\rm eff}^2}{\pi}\int_0^\infty{\rm d}x \frac{x^2}{1 + x^2}{\rm csch}^2\left(\frac12 g_{\rm eff}\sqrt{1 + x^2}\right)\,,
\end{align}
which tends to $1$ as $g_{\rm eff}\to 0$, and whose value at $g_{\rm eff}^2 = 0.4$ is 0.66.

Assuming the curvaton remains underdamped, the nonadiabatic correction to its damping rate is
\begin{align}
    \Upsilon = \Upsilon^{\rm ad}\frac{\Gamma_{\rm th}^2}{\Gamma_{\rm th}^2 + m_\sigma^2}\,.
\end{align}
The time at which the curvaton damps its energy into the plasma is then approximately set by
\begin{align}
    1 \approx \int{\rm dt}\Upsilon = \sqrt{\frac{90}{\pi^2g_\star}}\frac{\Mpl\Upsilon^{\rm ad}}{ T^2 }\frac{\Gamma_{\rm th}^2}{m_\sigma^2}\left(1 - \frac{\Gamma_{\rm th}}{m_\sigma}\arctan\frac{m_\sigma}{\Gamma_{\rm th}}\right)\,.
\end{align}
The damping temperature is then approximately
\begin{align}
    T_{\rm damp}\approx \frac{m_\sigma T}{\Gamma_{\rm th}}\left(\frac{1}{q} + \frac{\pi}{2}\arctan\left[\frac{2}{\pi}\left(\frac{3}{q}\right)^{1/3}\right]\right)^{-1}\,,
\end{align}
with
\begin{align}
    q = \sqrt{\frac{90}{\pi^2 g_\star}}\frac{\Mpl \Gamma_{\rm th}^3\Upsilon^{\rm ad}}{m_\sigma^3 T^2}\,.
\end{align}
When applying these expressions, one should evaluate $g_\star$ at damping.

Below electroweak symmetry breaking, the curvaton mixes with the Higgs through the interaction $\mu v_{\rm ew}\sigma h$. For $m_\sigma\ll m_h$, the Higgs responds adiabatically to curvaton oscillations, with $h\simeq-\mu v_{\rm ew}\sigma/m_h^2$. Since Standard Model fermion masses are proportional to the Higgs expectation value, they acquire a field dependence
\begin{align}
    m_f(\sigma)\simeq m_f\left(1-\frac{\mu\sigma}{m_h^2}\right)\,,
\end{align}
so that $m_f\partial_\sigma m_f\simeq-\mu m_f^2/m_h^2$. Applying the same relaxation-time formalism to a fermion species $f$ gives
\begin{align}
    \Upsilon_{\rm off}\sim\frac{\mu^2}{8\pi}\left(\frac{m_f}{m_h}\right)^4\frac{\Gamma_{\rm th}}{\Gamma_{\rm th}^2+m_\sigma^2}\,.
\end{align}
Since the damping is dominated by the heaviest species in equilibrium, we can approximate the heaviest species mass by $m_f\sim T$, recovering the scaling of \cref{eqn:Upsilon_off}. Higgs mixing also induces interactions with gluons through heavy-quark loops, which we expect to give the same temperature scaling, though the numerical coefficient is less certain.

\section{Higgs equilibration rate}
\label{app:equilibration}

The off-shell damping rate depends on the Higgs equilibration rate $\Gamma_{\rm th}$ at which the Higgs distribution relaxes toward thermal equilibrium, defined by
\begin{align}\label{eqn:equilibration_rate_definition}
    \delta\dot n_H=-\Gamma_{\rm th}\delta n_H\,,
\end{align}
where $\delta n_H = n_{H} - n_{H,\rm eq}$, with $n_H$ the Higgs number density and $n_{H,\rm eq}$ its equilibrium value. We do so by linearizing the number-changing collision terms around thermal equilibrium (see e.g.~\cite{Bodeker:2012gs,Bodeker:2019ajh}). 

We parameterize a small departure from chemical equilibrium by an effective chemical potential $\nu_H$,
\begin{align}
    f_H(p,T,\nu_H)=\frac{1}{e^{(E-\nu_H)/T}-1}\,,
\end{align}
where $E=\sqrt{p^2+m_H^2(T)}$.
At linear order, the shift in the Higgs number density relative to zero chemical potential can then be written
\begin{align}\label{eqn:susceptibility_definition}
    \delta n_H = \chi_H \nu_H\,,
\end{align}
where the susceptibility is defined as
\begin{align}
    \chi_H\equiv\left.\frac{\partial n_H}{\partial\nu_H}\right|_{\nu_H=0}=\frac{4}{T}\int\frac{d^3p}{(2\pi)^3}f_H^{\rm eq}(1+f_H^{\rm eq})\,.
\end{align}
The factor of four accounts for the Higgs degrees of freedom, and using $m_H^2(T)\approx0.4T^2$ in the Higgs dispersion relation we obtain $\chi_H\approx0.5T^2$. 

Consider a reaction $r$ that changes the number of Higgs by $\Delta N_{H,r}$, defined as the number in the final state minus that in the initial state. Detailed balance implies that the rate of the forward reaction $\gamma_{r,\rm forward}$ is related to the rate of the backward reaction $\gamma_{r,\rm backward}$ by
\begin{align}
    \frac{\gamma_{r,\rm forward}}{\gamma_{r,\rm backward}}=\exp\left(-\frac{\Delta N_{H,r}\nu_H}{T}\right)\,.
\end{align}
Expanding for $|\nu_H|\ll T$, and denoting the reaction rate at $\nu_{H} = 0$ by $\gamma_{r,\rm eq}$, gives
\begin{align}
    \gamma_{r,\rm forward}-\gamma_{r,\rm backward} = -\gamma_{r,\rm eq}\frac{\Delta N_{H,r}\nu_H}{T}\,.
\end{align}
Each net forward reaction changes the Higgs number by $\Delta N_{H,r}$, so the linearized collision term is
\begin{subequations}
\begin{align}
    \left.\dot n_H\right|_{\rm coll}&=\sum_r\Delta N_{H,r}(\gamma_{r,\rm forward}-\gamma_{r,\rm backward})\,,\\
    &=-\frac{\nu_H}{T}\sum_r(\Delta N_{H,r})^2\gamma_{r,\rm eq}\,.
\end{align}
\end{subequations}
For example, $HH^\dagger\leftrightarrow t\bar t$ changes the Higgs number by two, contributing $4\gamma_{r,\rm eq}$ to the sum. Comparing with \cref{eqn:equilibration_rate_definition}, and using \cref{eqn:susceptibility_definition} to eliminate $\nu_H$ gives
\begin{align}
    \Gamma_{\rm th}=\frac{1}{T\chi_H}\sum_r(\Delta N_{H,r})^2\gamma_{r,\rm eq}\,.
\end{align}
The remaining task is to estimate $\gamma_{r,\rm eq}$.

To estimate the equilibration rate, we retain the leading top-Yukawa--QCD reactions, represented by $t\bar t\leftrightarrow hg$ and $tg\leftrightarrow th$. These involve the large couplings $y_t$ and $g_3$ and provide a substantial contribution to Higgs equilibration~\cite{Kozaczuk:2015owa}. For an analytic estimate, we neglect external masses, use Maxwell--Boltzmann distributions in the reaction densities, and omit final-state occupation factors. The equilibrium reaction density is then~\cite{Gondolo:1990dk,Giudice:2003jh}
\begin{align}
    \gamma_{ab\to cd} &\approx \frac{g_ag_bT}{32\pi^4}\int_0^\infty {\rm d}s\,s^{3/2}K_1(\sqrt{s}/T)\sigma_{ab\to cd}(s)\,.
    \label{eq:gamma_r}
\end{align}
Here $s$ is the squared center-of-mass energy, $g_i$ counts internal degrees of freedom, and $K_1$ is a modified Bessel function. Cross sections are averaged over initial spins and colors and summed over final states. All incoming pairs considered here contain distinct species.

Soft fermion exchange produces logarithmic enhancements regulated by thermal self-energies. Approximating the regulator by $\Lambda_f^2=g_3^2T^2/6$, the matrix elements in Table~1 of Ref.~\cite{Kozaczuk:2015owa} give, for $s\gg\Lambda_f^2$,
\begin{align}
    \sigma_{t\bar t\to hg} &\approx \frac{y_t^2g_3^2}{\pi g_t^2 s}\left[\ln\frac{s}{\Lambda_f^2}-1\right]\,,\nonumber\\
    \sigma_{tg\to th} &\approx \frac{y_t^2g_3^2}{2\pi g_tg_g s}\ln\frac{s}{\Lambda_f^2}\,,
    \label{eq:sigma_tt_hg_log}
\end{align}
where $g_t=6$ and $g_g=16$. Extending these high-energy expressions through the thermal integral gives
\begin{align}
    \gamma_{t\bar t\to hg} &\approx \frac{y_t^2g_3^2T^4}{8\pi^5}\left[\ln\frac{4T^2}{\Lambda_f^2}-2\gamma_E\right]\,,\nonumber\\
    \gamma_{tg\to th} &\approx \frac{y_t^2g_3^2T^4}{16\pi^5}\left[\ln\frac{4T^2}{\Lambda_f^2}-2\gamma_E+1\right]\,,
    \label{eq:gamma_tt_hg_log}
\end{align}
for annihilation and scattering, respectively. Including the four Higgs degrees of freedom and charge-conjugate reactions gives four annihilation channels and eight scattering channels. Each changes the Higgs number by one, so
\begin{align}
    \Gamma_{\rm th}^{(t3)} &\approx \frac{4\gamma_{t\bar t\to hg}+8\gamma_{tg\to th}}{T\chi_H}\approx 0.018T\,,
    \label{eq:Gamma_H_all_classes}
\end{align}
where we have taken $y_t=0.95$, and $g_3=1.2$. As this estimate made a number of simplifications, we treat it as an order of magnitude estimate and use $\Gamma_{\rm th} = (0.01-0.05)T$ as a fiducial estimate of the true rate.

\section{Perturbative decay}
\label{app:decay}

When computing decay of the curvaton in-medium, we use the in-medium masses to set kinematic thresholds. In the symmetric phase, we include the effect of Bose enhancement for a decay to Higgs. After the EW symmetry breaking, we neglect the effect of Bose enhancement and Pauli blocking, and use the vacuum decay widths. 

The interaction in \cref{eqn:interaction} mediates curvaton decay into Higgs excitations. In the electroweak-symmetric phase, the tree-level rate is
\begin{align}
    \Gamma_{\sigma\to HH^\dagger}(T) = \frac{\mu^2}{8\pi m_\sigma}\sqrt{1-\frac{4m_H^2(T)}{m_\sigma^2}}\left[1 + 2 f_B\cbrac{\frac{m_\sigma}{2}}\right]\,,
\end{align}
when this channel is kinematically open. Here, $f_B = 1/(e^{E/T}-1)$ denotes the Bose distribution function. Away from threshold, the corresponding parametric width is
\begin{align}
    \Gamma_{\sigma\to HH^\dagger}
    \sim\frac{\mu^2}{8\pi m_\sigma}
    \simeq4\times10^{-17}\GeV\left(\frac{\mu}{1\keV}\right)^2\left(\frac{1\TeV}{m_\sigma}\right)\,.
    \label{eq:higgs_portal_width_scaling}
\end{align}

After electroweak symmetry breaking, the portal induces mixing between the curvaton and the physical Higgs boson. With $v=246\GeV$ denoting the Higgs vacuum expectation value, the mixing angle satisfies
\begin{align}\label{eqn:tan2theta}
    \tan(2\theta)=\frac{2\mu v}{m_\sigma^2-m_{h,0}^2}\,,
\end{align}
where $m_{h,0}$ is the Higgs mass after the EW symmetry breaking but before mixing. For small mixing away from $m_\sigma=m_{h,0}$,
\begin{align}
    \label{eq: mixing angle}
    \theta\sim\frac{\mu v}{m_\sigma^2-m_{h,0}^2}\,.
\end{align}
We denote fluctuations about the shifted vacuum by $S$, whose physical mass is $m_S$. The mixing-mediated partial width is
\begin{align}
    \label{eq: decay rate (after EWSB)}
    \Gamma_{S\to\mathrm{SM}}^{\mathrm{(mix)}}=\sin^2\theta\,\Gamma_h(m_S)\,,
\end{align}
where $\Gamma_h(m_S)$ is the width of a hypothetical Standard Model Higgs boson with mass $m_S$. When $m_S>2m_h$, the portal also permits direct decay into two Higgs bosons. In addition to the contribution from $\sigma H^\dagger H$ couplings, the SM quartic coupling $\lambda_H(H^\dagger H)^2$ also mediates this decay. To leading order in mixing, the effective trilinear coupling is expressed as
\begin{align}
    g_{Shh} \simeq \frac{m_S^2 + 2 m_h^2}{v} \sin\theta \cos^2\theta.
\end{align}
Correspondingly, the partial width is given by
\begin{align}
    \Gamma(S\to hh)
    &\simeq \frac{\mu^2}{32\pi m_S}
    \left(\frac{m_S^2+2m_h^2}{m_S^2-m_h^2}\right)^2
    \sqrt{1-\frac{4m_h^2}{m_S^2}}\,.
\end{align}

In the opposite mass hierarchy ($m_h > 2 m_S$), the Higgs is able to decay to two scalars. The corresponding trilinear coupling, which originates from the quartic interaction $\lambda_H(H^\dagger H)^2$ in the symmetric phase, is
\begin{align}
    g_{hSS}
    =\frac{m_h^2+2m_S^2}{v}
    \cos\theta\sin^2\theta\,,
\end{align}
leading to the partial decay width:
\begin{align}
    \Gamma(h\to SS)
    =\frac{g_{hSS}^2}{32\pi m_h}
    \sqrt{1-\frac{4m_S^2}{m_h^2}}\,.
\end{align}

Thus, at this order the total broken-phase width is
\begin{align}
    \Gamma_{S\to\mathrm{SM}}=\Gamma(S\to hh)+\sin^2\theta\,\Gamma_h(m_S)\,.
\end{align}
For the light masses of interest, the characteristic scaling of the vacuum decay width is
\begin{align}
    \Gamma_{\rm vac}\sim\frac{\mu^2m_\sigma^3}{8\pi m_h^4}\,,\qquad m_\sigma\sim\GeV\,.
\end{align}

The partial widths into fermions and massless gauge bosons are~\cite{Djouadi:2005gj,Krnjaic:2015mbs}
\begin{align}
    \Gamma(S\to f\bar f)
    &=N_c^f\sin^2\theta\,\frac{G_Fm_f^2m_S}{4\sqrt{2}\pi}\left(1-\frac{4m_f^2}{m_S^2}\right)^{3/2},\\
    \Gamma(S\to\gamma\gamma)
    &=\sin^2\theta\,\frac{G_F\alpha_\mathrm{EM}^2m_S^3}{128\sqrt{2}\pi^3}\nonumber\\
    &\quad\times\left|\sum_f N_c^fQ_f^2A_{1/2}(\tau_f)+A_1(\tau_W)\right|^2,\\
    \Gamma(S\to gg)
    &=\sin^2\theta\,\frac{G_F\alpha_s^2(m_S)m_S^3}{36\sqrt{2}\pi^3}\left|\frac34\sum_qA_{1/2}(\tau_q)\right|^2.
\end{align}
Here $\alpha_\mathrm{EM}\simeq1/137$ is the fine-structure constant, $\tau_i\equiv m_S^2/(4m_i^2)$, $N_c^f=3$ for quarks and $N_c^f=1$ for leptons, and $Q_f$ is the fermion electric charge. The loop functions are
\begin{align}
    A_{1/2}(\tau)&=\frac{2[\tau+(\tau-1)f(\tau)]}{\tau^2},\\
    A_1(\tau)&=-\frac{2\tau^2+3\tau+3(2\tau-1)f(\tau)}{\tau^2},
\end{align}
where
\begin{align}
    f(\tau)=\begin{cases}
        \arcsin^2\sqrt\tau, & \tau\leq1,\\
        -\dfrac14\left[\log\left(\dfrac{1+\sqrt{1-1/\tau}}{1-\sqrt{1-1/\tau}}\right)-i\pi\right]^2, & \tau>1.
    \end{cases}
\end{align}

Hadronic decay channels open when $m_S>2m_\pi$. Around $m_S\sim\GeV$, the width has substantial hadronic uncertainties because chiral perturbation theory is no longer uniformly accurate and perturbative QCD is not yet applicable. We therefore use the widths from Ref.~\cite{Winkler:2018qyg} in this region, connecting them to the perturbative expressions at lower and higher masses. In applying the results, we neglect the difference between the symmetric-phase quadratic mass and the light physical scalar mass only when the mixing-induced shift is small.

\section{Cosmological bounds}
\label{app:curvaton_cosmological_bounds}

In this appendix, we outline the specific steps and assumptions that are used to construct the contours in \cref{fig:results}. 

\paragraph{Upper boundary from $A_s$.}
For a given inflationary scale $H_\mathrm{inf}$, the observed amplitude of the scalar power spectrum determines the initial misalignment of the curvaton in terms of the transfer efficiency $R_\sigma$. Its definition (\cref{eq:curvaton_Rsigma}) gives
\begin{align}
  \Xi_{\rm dec}=\frac{4R_\sigma}{3(1-R_\sigma)}\,.
  \label{eq:Xi_R_boundary}
\end{align}
Assuming the quadratic potential and the sudden decay approximation, \cref{eq:curvaton_Xi_dec_As} can be written as
\begin{align}
  \Xi_{\rm dec}&\simeq \mathcal C(T_{\rm dec})R_\sigma^2\,,
  \label{eq:Xi_C_boundary}\\
  \mathcal C(T_{\rm dec})&\equiv \frac{4\Gamma^2(5/4)}{27\pi^3 A_s} \left(\frac{g_\star(T_{\rm dec})}{g_\star(T_{\rm osc})}\right)^{1/4} \left(\frac{H_\star}{\Mpl}\right)^2 \sqrt{\frac{m_\sigma}{H_{\rm rad}(T_\mathrm{dec})}}\,.
  \label{eq:C_boundary}
\end{align}
Combining \cref{eq:Xi_R_boundary} with \cref{eq:Xi_C_boundary} gives
\begin{align}
  R_\sigma^{\pm}=\frac12\left[1\pm
  \sqrt{1-\frac{16}{3\mathcal C}}\right]\,.
  \label{eq:Rsigma_branches}
\end{align}
We note that $\mathcal C\geq16/3$ for real solutions to exist. To obtain the upper green boundary in \cref{fig:results}, we scan over values of $R_\sigma$ consistent with the observed $A_s$ and $f_\mathrm{NL}$. For a given $R_\sigma$, we identify the corresponding initial curvaton misalignment and transfer temperature, and obtain $\mu$ which satisfies the integrated damping condition \cref{eqn:decay_time}, including both perturbative decay and off-shell damping. When evaluating this condition, we include the contribution of the curvaton to the expansion rate
\begin{align}
  H(T) = H_{\rm rad}(T)\sqrt{1+\Xi(T)}\,.
\end{align}
The upper green boundary in \cref{fig:results} is then obtained by maximizing $\mu$ over $R_\sigma$, subject to the perturbativity requirement as discussed below.
\paragraph{Perturbativity bound.}
We also require that the thermal Higgs mass remains large compared to the variation of Higgs mass due to the curvaton oscillation. Rewriting the condition in \cref{eqn:perturbativity} in terms of $\Xi$, 
\begin{align}
  \epsilon_H(T)&\equiv\frac{|\mu|\bar\sigma}{g_{\rm eff}^2T^2}=\sqrt{\frac{\pi^2g_\star}{15}}\sqrt{\Xi(T)}\frac{|\mu|}{g_{\rm eff}^2m_{\sigma}}\lesssim1\,.
  \label{eq:epsilon_boundary_appendix}
\end{align}
For the perturbativity line (green dot-dashed line in Fig.~\ref{fig:results}), we evaluate this condition at $T=\max(T_{\rm dec},T_{\rm ew})$ as this condition is derived by using the thermal Higgs mass estimated in the symmetric phase. Note that this neglects the strengthening of the constraint near the electroweak phase transition---we leave a detailed treatment of the crossover to future work. Parameter space is considered unconstrained if at least one $R_\sigma$ satisfies $\epsilon_H \lesssim 1$.
\paragraph{Lower boundary from BBN.}
The lower bound (orange line in Fig.~\ref{fig:results}) requires that energy transfer from the curvaton sector to the SM sector takes place early enough so that the resulting plasma has a temperature $T \gtrsim T_\mathrm{BBN} = 4\MeV$. Here, reheating refers to energy transfer of curvaton energy to the pre-existing SM thermal bath, rather than the one driven by the inflaton. We determine the reheating completion time $t_\mathrm{RH}$ as the moment which satisfies
\begin{align}
  \int_{t_{\rm osc}}^{t_{\rm RH}}\Upsilon_{\rm tot}(t)\,dt\geq1\,,\qquad\frac{\rho_\sigma(t_{\rm RH})}{\rho_r(t_{\rm RH})}\leq0.1\,.
\end{align}
We further require that the curvaton is subdominant until BBN. The reheating temperature is thus obtained from
\begin{align}
  \rho_r(t_{\rm RH})=\frac{\pi^2}{30}g_\star(T_{\rm RH})T_{\rm RH}^4\,,
\end{align}
and we impose $T_{\rm RH}\gtrsim T_{\rm BBN}$. It is possible that the curvaton comes to dominate after BBN, which may lead to further constraints, though we do not treat those here.

\bibliographystyle{utcaps_mod} 
\bibliography{bibliography}

@article{Cyncynates:2024ufu,
    author = "Cyncynates, David and Simon, Olivier",
    title = "{Scalar Relics from the Hot Big Bang}",
    eprint = "2410.22409",
    archivePrefix = "arXiv",
    primaryClass = "hep-ph",
    doi = "10.1103/wjt8-9dh9",
    journal = "Phys. Rev. Lett.",
    volume = "135",
    number = "10",
    pages = "101003",
    year = "2025"
}

@article{Enqvist:2013gwf,
    author = "Enqvist, Kari and Lerner, Rose N. and Takahashi, Tomo",
    title = "{The minimal curvaton-higgs model}",
    eprint = "1310.1374",
    archivePrefix = "arXiv",
    primaryClass = "astro-ph.CO",
    reportNumber = "HIP-2013-21-TH, DESY-13-180, DESY13-180",
    doi = "10.1088/1475-7516/2014/01/006",
    journal = "JCAP",
    volume = "01",
    pages = "006",
    year = "2014"
}

@article{Cyncynates:2024bxw,
    author = "Cyncynates, David and Simon, Olivier",
    title = "{Minimal targets for dilaton direct detection}",
    eprint = "2408.16816",
    archivePrefix = "arXiv",
    primaryClass = "hep-ph",
    doi = "10.1103/8mc9-6cmr",
    journal = "Phys. Rev. D",
    volume = "112",
    number = "5",
    pages = "055002",
    year = "2025"
}

@article{Lyth:2001nq,
    author = "Lyth, David H. and Wands, David",
    title = "{Generating the curvature perturbation without an inflaton}",
    eprint = "hep-ph/0110002",
    archivePrefix = "arXiv",
    reportNumber = "PU-RCG-01-33",
    doi = "10.1016/S0370-2693(01)01366-1",
    journal = "Phys. Lett. B",
    volume = "524",
    pages = "5--14",
    year = "2002"
}

@article{Planck:2019kim,
    author = "Akrami, Y. and others",
    collaboration = "Planck",
    title = "{Planck 2018 results. IX. Constraints on primordial non-Gaussianity}",
    eprint = "1905.05697",
    archivePrefix = "arXiv",
    primaryClass = "astro-ph.CO",
    doi = "10.1051/0004-6361/201935891",
    journal = "Astron. Astrophys.",
    volume = "641",
    pages = "A9",
    year = "2020"
}

@article{Planck:2018jri,
    author = "Akrami, Y. and others",
    collaboration = "Planck",
    title = "{Planck 2018 results. X. Constraints on inflation}",
    eprint = "1807.06211",
    archivePrefix = "arXiv",
    primaryClass = "astro-ph.CO",
    doi = "10.1051/0004-6361/201833887",
    journal = "Astron. Astrophys.",
    volume = "641",
    pages = "A10",
    year = "2020"
}

@article{Planck:2018vyg,
    author = "Aghanim, N. and others",
    collaboration = "Planck",
    title = "{Planck 2018 results. VI. Cosmological parameters}",
    eprint = "1807.06209",
    archivePrefix = "arXiv",
    primaryClass = "astro-ph.CO",
    doi = "10.1051/0004-6361/201833910",
    journal = "Astron. Astrophys.",
    volume = "641",
    pages = "A6",
    year = "2020",
    note = "[Erratum: Astron.Astrophys. 652, C4 (2021)]"
}

@article{Enqvist:2001zp,
    author = "Enqvist, Kari and Sloth, Martin S.",
    title = "{Adiabatic CMB perturbations in pre - big bang string cosmology}",
    eprint = "hep-ph/0109214",
    archivePrefix = "arXiv",
    reportNumber = "HIP-2001-51-TH",
    doi = "10.1016/S0550-3213(02)00043-3",
    journal = "Nucl. Phys. B",
    volume = "626",
    pages = "395--409",
    year = "2002"
}

@article{Bartolo:2002vf,
    author = "Bartolo, N and Liddle, Andrew R",
    title = "{The Simplest curvaton model}",
    eprint = "astro-ph/0203076",
    archivePrefix = "arXiv",
    doi = "10.1103/PhysRevD.65.121301",
    journal = "Phys. Rev. D",
    volume = "65",
    pages = "121301",
    year = "2002"
}

@article{Lyth:2002my,
    author = "Lyth, David H. and Ungarelli, Carlo and Wands, David",
    title = "{The Primordial density perturbation in the curvaton scenario}",
    eprint = "astro-ph/0208055",
    archivePrefix = "arXiv",
    reportNumber = "PU-ICG-02-15",
    doi = "10.1103/PhysRevD.67.023503",
    journal = "Phys. Rev. D",
    volume = "67",
    pages = "023503",
    year = "2003"
}

@article{Moroi:2001ct,
    author = "Moroi, Takeo and Takahashi, Tomo",
    title = "{Effects of cosmological moduli fields on cosmic microwave background}",
    eprint = "hep-ph/0110096",
    archivePrefix = "arXiv",
    reportNumber = "TU-632",
    doi = "10.1016/S0370-2693(01)01295-3",
    journal = "Phys. Lett. B",
    volume = "522",
    pages = "215--221",
    year = "2001",
    note = "[Erratum: Phys.Lett.B 539, 303--303 (2002)]"
}

@article{Krnjaic:2015mbs,
    author = "Krnjaic, Gordan",
    title = "{Probing Light Thermal Dark-Matter With a Higgs Portal Mediator}",
    eprint = "1512.04119",
    archivePrefix = "arXiv",
    primaryClass = "hep-ph",
    reportNumber = "FERMILAB-PUB-15-550-A",
    doi = "10.1103/PhysRevD.94.073009",
    journal = "Phys. Rev. D",
    volume = "94",
    number = "7",
    pages = "073009",
    year = "2016"
}

@article{Winkler:2018qyg,
    author = "Winkler, Martin Wolfgang",
    title = "{Decay and detection of a light scalar boson mixing with the Higgs boson}",
    eprint = "1809.01876",
    archivePrefix = "arXiv",
    primaryClass = "hep-ph",
    reportNumber = "NORDITA-2018-087",
    doi = "10.1103/PhysRevD.99.015018",
    journal = "Phys. Rev. D",
    volume = "99",
    number = "1",
    pages = "015018",
    year = "2019"
}

@article{Djouadi:2005gj,
    author = "Djouadi, Abdelhak",
    title = "{The Anatomy of electro-weak symmetry breaking. II. The Higgs bosons in the minimal supersymmetric model}",
    eprint = "hep-ph/0503173",
    archivePrefix = "arXiv",
    reportNumber = "LPT-ORSAY-05-18",
    doi = "10.1016/j.physrep.2007.10.005",
    journal = "Phys. Rept.",
    volume = "459",
    pages = "1--241",
    year = "2008"
}

@article{Ferber:2023iso,
    author = "Ferber, Torben and Grohsjean, Alexander and Kahlhoefer, Felix",
    title = "{Dark Higgs bosons at colliders}",
    eprint = "2305.16169",
    archivePrefix = "arXiv",
    primaryClass = "hep-ph",
    reportNumber = "P3H-23-034, TTP23-018",
    doi = "10.1016/j.ppnp.2024.104105",
    journal = "Prog. Part. Nucl. Phys.",
    volume = "136",
    pages = "104105",
    year = "2024"
}

@article{Fradette:2018hhl,
    author = "Fradette, Anthony and Pospelov, Maxim and Pradler, Josef and Ritz, Adam",
    title = "{Cosmological beam dump: constraints on dark scalars mixed with the Higgs boson}",
    eprint = "1812.07585",
    archivePrefix = "arXiv",
    primaryClass = "hep-ph",
    doi = "10.1103/PhysRevD.99.075004",
    journal = "Phys. Rev. D",
    volume = "99",
    number = "7",
    pages = "075004",
    year = "2019"
}

@article{Yu:2026xtb,
    author = "Yu, Yue and Maitra, Writasree and Dev, P. S. Bhupal and Fortin, Jean-Franccois and Harris, Steven P. and Sinha, Kuver and Zhang, Yongchao",
    title = "{Flash from the Past: New Gamma-Ray Constraints on Light CP-even Scalar from SN1987A}",
    eprint = "2603.08695",
    archivePrefix = "arXiv",
    primaryClass = "hep-ph",
    month = "3",
    year = "2026"
}

@article{Joseph:2026nut,
    author = "Joseph, Melissa and Liebersbach, Samuel and Madathil, Anirudhan A. and Marques-Tavares, Gustavo",
    title = "{Improved supernova bounds on CP-even scalars: cooling and decay constraints}",
    eprint = "2603.04513",
    archivePrefix = "arXiv",
    primaryClass = "hep-ph",
    month = "3",
    year = "2026"
}

@article{Balaji:2022noj,
    author = "Balaji, Shyam and Dev, P. S. Bhupal and Silk, Joseph and Zhang, Yongchao",
    title = "{Improved stellar limits on a light CP-even scalar}",
    eprint = "2205.01669",
    archivePrefix = "arXiv",
    primaryClass = "hep-ph",
    doi = "10.1088/1475-7516/2022/12/024",
    journal = "JCAP",
    volume = "12",
    pages = "024",
    year = "2022"
}

@article{Hardy:2024gwy,
    author = "Hardy, Edward and Sokolov, Anton and Stubbs, Henry",
    title = "{Supernova bounds on new scalars from resonant and soft emission}",
    eprint = "2410.17347",
    archivePrefix = "arXiv",
    primaryClass = "hep-ph",
    doi = "10.1007/JHEP04(2025)013",
    journal = "JHEP",
    volume = "04",
    pages = "013",
    year = "2025"
}

@article{ATLAS:2022vkf,
    author = "Aad, Georges and others",
    collaboration = "ATLAS",
    title = "{A detailed map of Higgs boson interactions by the ATLAS experiment ten years after the discovery}",
    eprint = "2207.00092",
    archivePrefix = "arXiv",
    primaryClass = "hep-ex",
    reportNumber = "CERN-EP-2022-057",
    doi = "10.1038/s41586-022-04893-w",
    journal = "Nature",
    volume = "607",
    number = "7917",
    pages = "52--59",
    year = "2022",
    note = "[Erratum: Nature 612, E24 (2022)]"
}

@article{CMS:2022dwd,
    author = "Tumasyan, Armen and others",
    collaboration = "CMS",
    title = "{A portrait of the Higgs boson by the CMS experiment ten years after the discovery.}",
    eprint = "2207.00043",
    archivePrefix = "arXiv",
    primaryClass = "hep-ex",
    reportNumber = "CMS-HIG-22-001, CERN-EP-2022-039",
    doi = "10.1038/s41586-022-04892-x",
    journal = "Nature",
    volume = "607",
    number = "7917",
    pages = "60--68",
    year = "2022",
    note = "[Erratum: Nature 623, (2023)]"
}

@article{LHCb:2015nkv,
    author = "Aaij, Roel and others",
    collaboration = "LHCb",
    title = "{Search for hidden-sector bosons in $B^0 \!\to K^{*0}\mu^+\mu^-$ decays}",
    eprint = "1508.04094",
    archivePrefix = "arXiv",
    primaryClass = "hep-ex",
    reportNumber = "CERN-PH-EP-2015-202, LHCB-PAPER-2015-036",
    doi = "10.1103/PhysRevLett.115.161802",
    journal = "Phys. Rev. Lett.",
    volume = "115",
    number = "16",
    pages = "161802",
    year = "2015"
}

@article{LHCb:2016awg,
    author = "Aaij, R. and others",
    collaboration = "LHCb",
    title = "{Search for long-lived scalar particles in $B^+ \to K^+ \chi (\mu^+\mu^-)$ decays}",
    eprint = "1612.07818",
    archivePrefix = "arXiv",
    primaryClass = "hep-ex",
    reportNumber = "CERN-EP-2016-302, LHCB-PAPER-2016-052",
    doi = "10.1103/PhysRevD.95.071101",
    journal = "Phys. Rev. D",
    volume = "95",
    number = "7",
    pages = "071101",
    year = "2017"
}

@article{NA62:2025upx,
    author = "Cortina Gil, Eduardo and others",
    collaboration = "NA62",
    title = "{Searches for hidden sectors using $K^+\to \pi^+X$ decays}",
    eprint = "2507.17286",
    archivePrefix = "arXiv",
    primaryClass = "hep-ex",
    reportNumber = "CERN-EP-2025-167",
    doi = "10.1007/JHEP11(2025)143",
    journal = "JHEP",
    volume = "11",
    pages = "143",
    year = "2025"
}

@article{CHARM:1985anb,
    author = "Bergsma, F. and others",
    collaboration = "CHARM",
    title = "{Search for Axion Like Particle Production in 400-{GeV} Proton - Copper Interactions}",
    reportNumber = "CERN-EP-85-38",
    doi = "10.1016/0370-2693(85)90400-9",
    journal = "Phys. Lett. B",
    volume = "157",
    pages = "458--462",
    year = "1985"
}

@article{NA62:2025yzs,
    author = "Cortina Gil, Eduardo and others",
    collaboration = "NA62",
    title = "{Search for hadronic decays of feebly-interacting particles at NA62}",
    eprint = "2502.04241",
    archivePrefix = "arXiv",
    primaryClass = "hep-ex",
    reportNumber = "CERN-EP-2025-012",
    doi = "10.1140/epjc/s10052-025-14133-w",
    journal = "Eur. Phys. J. C",
    volume = "85",
    number = "5",
    pages = "571",
    year = "2025"
}

@article{Alekhin:2015byh,
    author = "Alekhin, Sergey and others",
    title = "{A facility to Search for Hidden Particles at the CERN SPS: the SHiP physics case}",
    eprint = "1504.04855",
    archivePrefix = "arXiv",
    primaryClass = "hep-ph",
    reportNumber = "CERN-SPSC-2015-017, SPSC-P-350-ADD-1",
    doi = "10.1088/0034-4885/79/12/124201",
    journal = "Rept. Prog. Phys.",
    volume = "79",
    number = "12",
    pages = "124201",
    year = "2016"
}

@article{Curtin:2018mvb,
    author = "Curtin, David and others",
    title = "{Long-Lived Particles at the Energy Frontier: The MATHUSLA Physics Case}",
    eprint = "1806.07396",
    archivePrefix = "arXiv",
    primaryClass = "hep-ph",
    reportNumber = "FERMILAB-PUB-18-264-T",
    doi = "10.1088/1361-6633/ab28d6",
    journal = "Rept. Prog. Phys.",
    volume = "82",
    number = "11",
    pages = "116201",
    year = "2019"
}

@article{Fradette:2017sdd,
    author = "Fradette, Anthony and Pospelov, Maxim",
    title = "{BBN for the LHC: constraints on lifetimes of the Higgs portal scalars}",
    eprint = "1706.01920",
    archivePrefix = "arXiv",
    primaryClass = "hep-ph",
    doi = "10.1103/PhysRevD.96.075033",
    journal = "Phys. Rev. D",
    volume = "96",
    number = "7",
    pages = "075033",
    year = "2017"
}

@article{Karagiannis:2019jjx,
    author = "Karagiannis, Dionysios and Slosar, An{\v{z}}e and Liguori, Michele",
    title = "{Forecasts on Primordial non-Gaussianity from 21 cm Intensity Mapping experiments}",
    eprint = "1911.03964",
    archivePrefix = "arXiv",
    primaryClass = "astro-ph.CO",
    doi = "10.1088/1475-7516/2020/11/052",
    journal = "JCAP",
    volume = "11",
    pages = "052",
    year = "2020"
}

@article{Heinrich:2023qaa,
    author = "Heinrich, Chen and Dore, Olivier and Krause, Elisabeth",
    title = "{Measuring fNL with the SPHEREx multitracer redshift space bispectrum}",
    eprint = "2311.13082",
    archivePrefix = "arXiv",
    primaryClass = "astro-ph.CO",
    doi = "10.1103/PhysRevD.109.123511",
    journal = "Phys. Rev. D",
    volume = "109",
    number = "12",
    pages = "123511",
    year = "2024"
}

@article{Euclid:2024ris,
    author = "Andrews, A. and others",
    collaboration = "Euclid",
    title = "{Euclid: Field-level inference of primordial non-Gaussianity and cosmic initial conditions}",
    eprint = "2412.11945",
    archivePrefix = "arXiv",
    primaryClass = "astro-ph.CO",
    month = "12",
    year = "2024"
}

@article{Starobinsky:1980te,
    author = "Starobinsky, Alexei A.",
    title = "{A New Type of Isotropic Cosmological Models Without Singularity}",
    doi = "10.1016/0370-2693(80)90670-X",
    journal = "Phys. Lett. B",
    volume = "91",
    pages = "99--102",
    year = "1980"
}

@article{Guth:1980zm,
    author = "Guth, Alan H.",
    title = "{The Inflationary Universe: A Possible Solution to the Horizon and Flatness Problems}",
    doi = "10.1103/PhysRevD.23.347",
    journal = "Phys. Rev. D",
    volume = "23",
    pages = "347--356",
    year = "1981"
}

@article{Linde:1981mu,
    author = "Linde, Andrei D.",
    title = "{A New Inflationary Universe Scenario: A Possible Solution of the Horizon, Flatness, Homogeneity, Isotropy and Primordial Monopole Problems}",
    doi = "10.1016/0370-2693(82)91219-9",
    journal = "Phys. Lett. B",
    volume = "108",
    pages = "389--393",
    year = "1982"
}

@article{Mukhanov:1981xt,
    author = "Mukhanov, Viatcheslav F. and Chibisov, G. V.",
    title = "{Quantum Fluctuations and a Nonsingular Universe}",
    journal = "JETP Lett.",
    volume = "33",
    pages = "532--535",
    year = "1981"
}

@article{Batell:2026sml,
    author = "Batell, Brian and Ghalsasi, Akshay and Ghosh, Subhajit and Rai, Mudit",
    title = "{Phasing out Dark Matter Isocurvature with Thermal Misalignment}",
    eprint = "2603.18132",
    archivePrefix = "arXiv",
    primaryClass = "hep-ph",
    reportNumber = "PITT-PACC-2605, UT-WI-09-2026, MI-HET-881",
    month = "3",
    year = "2026"
}

@article{Dolan:1973qd,
    author = "Dolan, L. and Jackiw, R.",
    title = "{Symmetry Behavior at Finite Temperature}",
    journal = "Phys. Rev. D",
    volume = "9",
    pages = "3320--3341",
    year = "1974",
    doi = "10.1103/PhysRevD.9.3320"
}

@article{Weinberg:1974hy,
    author = "Weinberg, Steven",
    title = "{Gauge and Global Symmetries at High Temperature}",
    journal = "Phys. Rev. D",
    volume = "9",
    pages = "3357--3378",
    year = "1974",
    doi = "10.1103/PhysRevD.9.3357"
}

@article{Batell:2021xsi,
    author = "Batell, Brian and Ghalsasi, Akshay",
    title = "{Thermal Misalignment of Scalar Dark Matter}",
    eprint = "2109.04476",
    archivePrefix = "arXiv",
    primaryClass = "hep-ph",
    journal = "Phys. Rev. D",
    volume = "107",
    number = "9",
    pages = "L091701",
    year = "2023",
    doi = "10.1103/PhysRevD.107.L091701"
}

@article{Batell:2022hsf,
    author = "Batell, Brian and Ghalsasi, Akshay and Rai, Mudit",
    title = "{Dynamics of Dark Matter Misalignment Through the Higgs Portal}",
    eprint = "2211.09132",
    archivePrefix = "arXiv",
    primaryClass = "hep-ph",
    reportNumber = "PITT-PACC-2213",
    journal = "JHEP",
    volume = "01",
    pages = "038",
    year = "2024",
    doi = "10.1007/JHEP01(2024)038"
}

@article{Curtin:2016urg,
    author = "Curtin, David and Meade, Patrick and Ramani, Harikrishnan",
    title = "{Thermal Resummation and Phase Transitions}",
    eprint = "1612.00466",
    archivePrefix = "arXiv",
    primaryClass = "hep-ph",
    reportNumber = "YITP-2016-48",
    doi = "10.1140/epjc/s10052-018-6268-0",
    journal = "Eur. Phys. J. C",
    volume = "78",
    number = "9",
    pages = "787",
    year = "2018"
}

@article{Arnold:1992rz,
    author = "Arnold, Peter Brockway and Espinosa, Olivier",
    title = "{The Effective potential and first order phase transitions: Beyond leading-order}",
    eprint = "hep-ph/9212235",
    archivePrefix = "arXiv",
    reportNumber = "UW-PT-92-18, USM-TH-60",
    doi = "10.1103/PhysRevD.47.3546",
    journal = "Phys. Rev. D",
    volume = "47",
    pages = "3546",
    year = "1993",
    note = "[Erratum: Phys.Rev.D 50, 6662 (1994)]"
}

@inproceedings{Quiros:1992ez,
    author = "Quiros, Mariano",
    title = "{On daisy and superdaisy resummation of the effective potential at finite temperature}",
    booktitle = "{4th Hellenic School on Elementary Particle Physics}",
    eprint = "hep-ph/9304284",
    archivePrefix = "arXiv",
    reportNumber = "IEM-FT-71-93",
    pages = "502--511",
    month = "9",
    year = "1992"
}

@article{Coleman:1973jx,
    author = "Coleman, Sidney R. and Weinberg, Erick J.",
    title = "{Radiative Corrections as the Origin of Spontaneous Symmetry Breaking}",
    doi = "10.1103/PhysRevD.7.1888",
    journal = "Phys. Rev. D",
    volume = "7",
    pages = "1888--1910",
    year = "1973"
}

@inproceedings{Quiros:1999jp,
    author = "Quiros, Mariano",
    title = "{Finite temperature field theory and phase transitions}",
    booktitle = "{ICTP Summer School in High-Energy Physics and Cosmology}",
    eprint = "hep-ph/9901312",
    archivePrefix = "arXiv",
    reportNumber = "IEM-FT-187-99",
    pages = "187--259",
    month = "1",
    year = "1999"
}

@article{SimonsObservatory:2025avm,
    author = "Abril-Cabezas, I. and others",
    collaboration = "Simons Observatory",
    title = "{The Simons Observatory: forecasted constraints on primordial gravitational waves with the expanded array of Small Aperture Telescopes}",
    eprint = "2512.15833",
    archivePrefix = "arXiv",
    primaryClass = "astro-ph.CO",
    reportNumber = "FERMILAB-PUB-25-0947-PPD",
    doi = "10.1088/1475-7516/2026/04/051",
    journal = "JCAP",
    volume = "04",
    pages = "051",
    year = "2026"
}

@article{Wands:2000dp,
    author = "Wands, David and Malik, Karim A. and Lyth, David H. and Liddle, Andrew R.",
    title = "{A New approach to the evolution of cosmological perturbations on large scales}",
    eprint = "astro-ph/0003278",
    archivePrefix = "arXiv",
    doi = "10.1103/PhysRevD.62.043527",
    journal = "Phys. Rev. D",
    volume = "62",
    pages = "043527",
    year = "2000"
}

@article{Rigopoulos:2003ak,
    author = "Rigopoulos, G. I. and Shellard, E. P. S.",
    title = "{The separate universe approach and the evolution of nonlinear superhorizon cosmological perturbations}",
    eprint = "astro-ph/0306620",
    archivePrefix = "arXiv",
    doi = "10.1103/PhysRevD.68.123518",
    journal = "Phys. Rev. D",
    volume = "68",
    pages = "123518",
    year = "2003"
}

@article{Salopek:1990jq,
    author = "Salopek, D. S. and Bond, J. R.",
    title = "{Nonlinear Evolution of Long Wavelength Metric Fluctuations in Inflationary Models}",
    reportNumber = "FERMILAB-PUB-90-131-A",
    doi = "10.1103/PhysRevD.42.3936",
    journal = "Phys. Rev. D",
    volume = "42",
    pages = "3936--3962",
    year = "1990"
}

@article{Berryman:2019dme,
    author = "Berryman, Jeffrey M. and de Gouvea, Andre and Fox, Patrick J and Kayser, Boris Jules and Kelly, Kevin James and Raaf, Jennifer Lynne",
    title = "{Searches for Decays of New Particles in the DUNE Multi-Purpose Near Detector}",
    eprint = "1912.07622",
    archivePrefix = "arXiv",
    primaryClass = "hep-ph",
    reportNumber = "FERMILAB-PUB-19-607-ND-T, NUHEP-TH/19-16",
    doi = "10.1007/JHEP02(2020)174",
    journal = "JHEP",
    volume = "02",
    pages = "174",
    year = "2020"
}

@article{HIKE:2023ext,
    author = "Ashraf, M. U. and others",
    collaboration = "HIKE",
    title = "{High Intensity Kaon Experiments (HIKE) at the CERN SPS Proposal for Phases 1 and 2}",
    eprint = "2311.08231",
    archivePrefix = "arXiv",
    primaryClass = "hep-ex",
    reportNumber = "CERN-SPSC-2023-031",
    month = "11",
    year = "2023"
}

@article{MicroBooNE:2022ctm,
    author = "Abratenko, P. and others",
    collaboration = "MicroBooNE",
    title = "{Search for long-lived heavy neutral leptons and Higgs portal scalars decaying in the MicroBooNE detector}",
    eprint = "2207.03840",
    archivePrefix = "arXiv",
    primaryClass = "hep-ex",
    reportNumber = "FERMILAB-PUB-22-507",
    doi = "10.1103/PhysRevD.106.092006",
    journal = "Phys. Rev. D",
    volume = "106",
    number = "9",
    pages = "092006",
    year = "2022"
}

@article{ATLAS:2023oaq,
    author = "Aad, Georges and others",
    collaboration = "ATLAS",
    title = "{Combined Measurement of the Higgs Boson Mass from the H{\textrightarrow}{\ensuremath{\gamma}}{\ensuremath{\gamma}} and H{\textrightarrow}ZZ*{\textrightarrow}4{\ensuremath{\ell}} Decay Channels with the ATLAS Detector Using s=7, 8, and 13~TeV pp Collision Data}",
    eprint = "2308.04775",
    archivePrefix = "arXiv",
    primaryClass = "hep-ex",
    reportNumber = "CERN-EP-2023-156",
    doi = "10.1103/PhysRevLett.131.251802",
    journal = "Phys. Rev. Lett.",
    volume = "131",
    number = "25",
    pages = "251802",
    year = "2023"
}

@inproceedings{Wang:2019mzd,
    author = "Wang, Yan and List, Jenny and Berggren, Mikael",
    collaboration = "International Large Detector concept group",
    title = "{Search for Extra Scalars Produced in Association with Muon Pairs at the ILC}",
    booktitle = "{International Workshop on Future Linear Colliders}",
    eprint = "1902.06118",
    archivePrefix = "arXiv",
    primaryClass = "hep-ex",
    month = "2",
    year = "2019"
}

@article{EuropeanStrategyforParticlePhysicsPreparatoryGroup:2019qin,
    author = "Ellis, Richard Keith and others",
    title = "{Physics Briefing Book}: {Input for the European Strategy for Particle Physics Update 2020}",
    eprint = "1910.11775",
    archivePrefix = "arXiv",
    primaryClass = "hep-ex",
    reportNumber = "CERN-ESU-004",
    month = "10",
    year = "2019"
}

@article{Bahl:2022igd,
    author = {Bahl, Henning and Biek{\"o}tter, Thomas and Heinemeyer, Sven and Li, Cheng and Paasch, Steven and Weiglein, Georg and Wittbrodt, Jonas},
    title = "{HiggsTools: BSM scalar phenomenology with new versions of HiggsBounds and HiggsSignals}",
    eprint = "2210.09332",
    archivePrefix = "arXiv",
    primaryClass = "hep-ph",
    doi = "10.1016/j.cpc.2023.108803",
    journal = "Comput. Phys. Commun.",
    volume = "291",
    pages = "108803",
    year = "2023"
}

@inproceedings{Batell:2022dpx,
    author = "Batell, Brian and Blinov, Nikita and Hearty, Christopher and McGehee, Robert",
    title = "{Exploring Dark Sector Portals with High Intensity Experiments}",
    booktitle = "{Snowmass 2021}",
    eprint = "2207.06905",
    archivePrefix = "arXiv",
    primaryClass = "hep-ph",
    month = "7",
    year = "2022"
}

@article{deBlas:2025gyz,
    author = "de Blas, Jorge and others",
    title = "{Physics Briefing Book: Input for the 2026 update of the European Strategy for Particle Physics}",
    eprint = "2511.03883",
    archivePrefix = "arXiv",
    primaryClass = "hep-ex",
    reportNumber = "CERN--2025-008, CERN-ESU-2025-001",
    doi = "10.23731/CYRM-2025-008",
    month = "11",
    year = "2025"
}

@inproceedings{BICEPKeck:2024stm,
    author = "Ade, P. A. R. and others",
    collaboration = "BICEP/Keck",
    title = "{Constraining Inflation with the BICEP/Keck CMB Polarization Experiments}",
    booktitle = "{58th Rencontres de Moriond on Cosmology}",
    eprint = "2405.19469",
    archivePrefix = "arXiv",
    primaryClass = "astro-ph.CO",
    doi = "10.58027/a73n-pn49",
    month = "5",
    year = "2024"
}

@article{Buchmuller:2004tu,
    author = "Buchmuller, W. and Di Bari, P. and Plumacher, M.",
    title = "{Some aspects of thermal leptogenesis}",
    eprint = "hep-ph/0406014",
    archivePrefix = "arXiv",
    reportNumber = "DESY-04-093",
    doi = "10.1088/1367-2630/6/1/105",
    journal = "New J. Phys.",
    volume = "6",
    pages = "105",
    year = "2004"
}

@article{Davidson:2002qv,
    author = "Davidson, Sacha and Ibarra, Alejandro",
    title = "{A Lower bound on the right-handed neutrino mass from leptogenesis}",
    eprint = "hep-ph/0202239",
    archivePrefix = "arXiv",
    reportNumber = "OUTP-02-10P, IPPP-02-16, DCPT-02-32",
    doi = "10.1016/S0370-2693(02)01735-5",
    journal = "Phys. Lett. B",
    volume = "535",
    pages = "25--32",
    year = "2002"
}

@article{Giudice:2003jh,
    author = "Giudice, G. F. and Notari, A. and Raidal, M. and Riotto, A. and Strumia, A.",
    title = "{Towards a complete theory of thermal leptogenesis in the SM and MSSM}",
    eprint = "hep-ph/0310123",
    archivePrefix = "arXiv",
    reportNumber = "IFUP-TH-2003-37, CERN-TH-2003-240, IFUP-TH/2003-37 and CERN-TH/2003-240",
    doi = "10.1016/j.nuclphysb.2004.02.019",
    journal = "Nucl. Phys. B",
    volume = "685",
    pages = "89--149",
    year = "2004"
}

@article{Bodeker:2019ajh,
    author = {B{\"o}deker, Dietrich and Schr{\"o}der, Dennis},
    title = "{Equilibration of right-handed electrons}",
    eprint = "1902.07220",
    archivePrefix = "arXiv",
    primaryClass = "hep-ph",
    doi = "10.1088/1475-7516/2019/05/010",
    journal = "JCAP",
    volume = "05",
    pages = "010",
    year = "2019"
}

@article{Bodeker:2012gs,
    author = "Bodeker, D. and Laine, M.",
    title = "{Heavy quark chemical equilibration rate as a transport coefficient}",
    eprint = "1205.4987",
    archivePrefix = "arXiv",
    primaryClass = "hep-ph",
    reportNumber = "BI-TP-2012-15, INT-PUB-12-026",
    doi = "10.1007/JHEP07(2012)130",
    journal = "JHEP",
    volume = "07",
    pages = "130",
    year = "2012"
}

@article{Gondolo:1990dk,
    author = "Gondolo, Paolo and Gelmini, Graciela",
    title = "{Cosmic abundances of stable particles: Improved analysis}",
    reportNumber = "UCLA-90-TEP-68",
    doi = "10.1016/0550-3213(91)90438-4",
    journal = "Nucl. Phys. B",
    volume = "360",
    pages = "145--179",
    year = "1991"
}

@article{Kozaczuk:2015owa,
    author = "Kozaczuk, Jonathan",
    title = "{Bubble Expansion and the Viability of Singlet-Driven Electroweak Baryogenesis}",
    eprint = "1506.04741",
    archivePrefix = "arXiv",
    primaryClass = "hep-ph",
    doi = "10.1007/JHEP10(2015)135",
    journal = "JHEP",
    volume = "10",
    pages = "135",
    year = "2015"
}

@article{MicroBooNE:2025jkl,
    author = "Abratenko, P. and others",
    collaboration = "MicroBooNE",
    title = "{Search for the production of Higgs-portal scalar bosons in the NuMI beam using the MicroBooNE detector}",
    eprint = "2501.08052",
    archivePrefix = "arXiv",
    primaryClass = "hep-ex",
    reportNumber = "FERMILAB-PUB-25-0012-PPD",
    doi = "10.1103/6myb-c3g1",
    journal = "Phys. Rev. D",
    volume = "113",
    number = "7",
    pages = "075007--6",
    year = "2026"
}

@article{ICARUS:2024oqb,
    author = "Alrahman, F. Abd and others",
    collaboration = "ICARUS",
    title = "{Search for a Hidden Sector Scalar from Kaon Decay in the Dimuon Final State at ICARUS}",
    eprint = "2411.02727",
    archivePrefix = "arXiv",
    primaryClass = "hep-ex",
    reportNumber = "FERMILAB-PUB-24-0581-PPD",
    doi = "10.1103/PhysRevLett.134.151801",
    journal = "Phys. Rev. Lett.",
    volume = "134",
    number = "15",
    pages = "151801",
    year = "2025"
}

@article{Ireland:2024wwn,
    author = "Ireland, Aurora and Krnjaic, Gordan and Okawa, Takuya",
    title = "{Affleck-Dine curvaton}",
    eprint = "2410.13712",
    archivePrefix = "arXiv",
    primaryClass = "hep-ph",
    reportNumber = "FERMILAB-PUB-24-0744-T",
    doi = "10.1103/7hr7-rtvh",
    journal = "Phys. Rev. D",
    volume = "113",
    number = "10",
    pages = "103537",
    year = "2026"
}

@article{Enqvist:2003mr,
    author = "Enqvist, Kari and Jokinen, Asko and Kasuya, Shinta and Mazumdar, Anupam",
    title = "{MSSM flat direction as a curvaton}",
    eprint = "hep-ph/0303165",
    archivePrefix = "arXiv",
    reportNumber = "HIP-2003-13-TH",
    doi = "10.1103/PhysRevD.68.103507",
    journal = "Phys. Rev. D",
    volume = "68",
    pages = "103507",
    year = "2003"
}

@article{Dimopoulos:2003az,
    author = "Dimopoulos, Konstantinos and Lyth, D. H. and Notari, A. and Riotto, A.",
    title = "{The Curvaton as a pseudoNambu-Goldstone boson}",
    eprint = "hep-ph/0304050",
    archivePrefix = "arXiv",
    reportNumber = "DFPD-TH-03-12",
    doi = "10.1088/1126-6708/2003/07/053",
    journal = "JHEP",
    volume = "07",
    pages = "053",
    year = "2003"
}

@article{Enqvist:2012tc,
    author = "Enqvist, Kari and Figueroa, Daniel G. and Lerner, Rose N.",
    title = "{Curvaton Decay by Resonant Production of the Standard Model Higgs}",
    eprint = "1211.5028",
    archivePrefix = "arXiv",
    primaryClass = "astro-ph.CO",
    reportNumber = "HIP-2012-19-TH",
    doi = "10.1088/1475-7516/2013/01/040",
    journal = "JCAP",
    volume = "01",
    pages = "040",
    year = "2013"
}

@article{Enqvist:2013qba,
    author = "Enqvist, Kari and Lerner, Rose N. and Rusak, Stanislav",
    title = "{Reheating dynamics affects non-perturbative decay of spectator fields}",
    eprint = "1308.3321",
    archivePrefix = "arXiv",
    primaryClass = "astro-ph.CO",
    reportNumber = "HIP-2013-11-TH",
    doi = "10.1088/1475-7516/2013/11/034",
    journal = "JCAP",
    volume = "11",
    pages = "034",
    year = "2013"
}

@article{Mukaida:2014yia,
    author = "Mukaida, Kyohei and Nakayama, Kazunori and Takimoto, Masahiro",
    title = "{Curvaton Dynamics Revisited}",
    eprint = "1401.5821",
    archivePrefix = "arXiv",
    primaryClass = "hep-ph",
    reportNumber = "UT-14-02",
    doi = "10.1088/1475-7516/2014/06/013",
    journal = "JCAP",
    volume = "06",
    pages = "013",
    year = "2014"
}

@article{Kitajima:2014xna,
    author = "Kitajima, Naoya and Langlois, David and Takahashi, Tomo and Takesako, Tomohiro and Yokoyama, Shuichiro",
    title = "{Thermal Effects and Sudden Decay Approximation in the Curvaton Scenario}",
    eprint = "1407.5148",
    archivePrefix = "arXiv",
    primaryClass = "astro-ph.CO",
    reportNumber = "ICRR-REPORT-690-2014-16",
    doi = "10.1088/1475-7516/2014/10/032",
    journal = "JCAP",
    volume = "10",
    pages = "032",
    year = "2014"
}

@inproceedings{Alimena:2025kjv,
    author = "Alimena, J. and others",
    title = "{Feebly-Interacting Particles: FIPs at LHCb {\textemdash} Workshop Report 2025 Edition}",
    booktitle = "{LHCb FIP Physics Workshop 2025}",
    eprint = "2510.05257",
    archivePrefix = "arXiv",
    primaryClass = "hep-ph",
    month = "10",
    year = "2025"
}

@article{Dimopoulos:2003ii,
    author = "Dimopoulos, Konstantinos and Lazarides, George and Lyth, David and Ruiz de Austri, Roberto",
    title = "{The Peccei-Quinn field as curvaton}",
    eprint = "hep-ph/0303154",
    archivePrefix = "arXiv",
    reportNumber = "UT-STPD-1-03",
    doi = "10.1088/1126-6708/2003/05/057",
    journal = "JHEP",
    volume = "05",
    pages = "057",
    year = "2003"
}

@article{Chun:2004gx,
    author = "Chun, Eung Jin and Dimopoulos, Konstantinos and Lyth, David",
    title = "{Curvaton and QCD axion in supersymmetric theories}",
    eprint = "hep-ph/0402059",
    archivePrefix = "arXiv",
    doi = "10.1103/PhysRevD.70.103510",
    journal = "Phys. Rev. D",
    volume = "70",
    pages = "103510",
    year = "2004"
}

@article{Mazumdar:2003bs,
    author = "Mazumdar, Anupam and Perez-Lorenzana, Abdel",
    title = "{Sneutrino condensate source for density perturbations, leptogenesis and low reheat temperature}",
    eprint = "hep-ph/0311106",
    archivePrefix = "arXiv",
    doi = "10.1103/PhysRevLett.92.251301",
    journal = "Phys. Rev. Lett.",
    volume = "92",
    pages = "251301",
    year = "2004"
}

@article{Piazza:2010ye,
    author = "Piazza, Federico and Pospelov, Maxim",
    title = "{Sub-eV scalar dark matter through the super-renormalizable Higgs portal}",
    eprint = "1003.2313",
    archivePrefix = "arXiv",
    primaryClass = "hep-ph",
    doi = "10.1103/PhysRevD.82.043533",
    journal = "Phys. Rev. D",
    volume = "82",
    pages = "043533",
    year = "2010"
}

@article{Banerjee:2025qyg,
    author = "Banerjee, Abhishek and Nguyen, Ngan H. and Tanin, Erwin H.",
    title = "{Thermal damping of mass-modulating scalars}",
    eprint = "2512.05188",
    archivePrefix = "arXiv",
    primaryClass = "hep-ph",
    doi = "10.1103/rqq9-1dc2",
    journal = "Phys. Rev. D",
    volume = "114",
    number = "1",
    pages = "015032",
    year = "2026"
}

@article{Bodeker:2022ihg,
    author = {B{\"o}deker, Dietrich and Nienaber, Jan},
    title = "{Scalar field damping at high temperatures}",
    eprint = "2205.14166",
    archivePrefix = "arXiv",
    primaryClass = "hep-ph",
    doi = "10.1103/PhysRevD.106.056016",
    journal = "Phys. Rev. D",
    volume = "106",
    number = "5",
    pages = "056016",
    year = "2022"
}

@article{NA62:2020pwi,
    author = "Cortina Gil, Eduardo and others",
    collaboration = "NA62",
    title = "{Search for $\pi^0$ decays to invisible particles}",
    eprint = "2010.07644",
    archivePrefix = "arXiv",
    primaryClass = "hep-ex",
    reportNumber = "CERN-EP-2020-193",
    doi = "10.1007/JHEP02(2021)201",
    journal = "JHEP",
    volume = "02",
    pages = "201",
    year = "2021"
}

@article{NA62:2020xlg,
    author = "Cortina Gil, Eduardo and others",
    collaboration = "NA62",
    title = "{Search for a feebly interacting particle  $X$ in the decay $K^{+}\rightarrow\pi^{+}X$}",
    eprint = "2011.11329",
    archivePrefix = "arXiv",
    primaryClass = "hep-ex",
    reportNumber = "CERN-EP-2020-227",
    doi = "10.1007/JHEP03(2021)058",
    journal = "JHEP",
    volume = "03",
    pages = "058",
    year = "2021"
}

@article{NA62:2021zjw,
    author = "Cortina Gil, Eduardo and others",
    collaboration = "NA62",
    title = "{Measurement of the very rare K$^{+}${\textrightarrow}$ {\pi}^{+}\nu \overline{\nu} $ decay}",
    eprint = "2103.15389",
    archivePrefix = "arXiv",
    primaryClass = "hep-ex",
    doi = "10.1007/JHEP06(2021)093",
    journal = "JHEP",
    volume = "06",
    pages = "093",
    year = "2021"
}

@article{Belle-II:2023ueh,
    author = "Adachi, I. and others",
    collaboration = "Belle-II",
    title = "{Search for a long-lived spin-0 mediator in b{\textrightarrow}s transitions at the Belle II experiment}",
    eprint = "2306.02830",
    archivePrefix = "arXiv",
    primaryClass = "hep-ex",
    reportNumber = "Belle II Preprint 2023-009, KEK Preprint 2023-7",
    doi = "10.1103/PhysRevD.108.L111104",
    journal = "Phys. Rev. D",
    volume = "108",
    number = "11",
    pages = "L111104",
    year = "2023"
}

@article{Gorbunov:2021ccu,
    author = "Gorbunov, Dmitry and Krasnov, Igor and Suvorov, Sergey",
    title = "{Constraints on light scalars from PS191 results}",
    eprint = "2105.11102",
    archivePrefix = "arXiv",
    primaryClass = "hep-ph",
    reportNumber = "INR-TH-2021-012",
    doi = "10.1016/j.physletb.2021.136524",
    journal = "Phys. Lett. B",
    volume = "820",
    pages = "136524",
    year = "2021"
}

@article{Foroughi-Abari:2020gju,
    author = "Foroughi-Abari, Saeid and Ritz, Adam",
    title = "{LSND Constraints on the Higgs Portal}",
    eprint = "2004.14515",
    archivePrefix = "arXiv",
    primaryClass = "hep-ph",
    doi = "10.1103/PhysRevD.102.035015",
    journal = "Phys. Rev. D",
    volume = "102",
    number = "3",
    pages = "035015",
    year = "2020"
}

@article{Ahdida:2023okr,
    author = "Ahdida, C. and others",
    title = "{Post-LS3 Experimental Options in ECN3}",
    eprint = "2310.17726",
    archivePrefix = "arXiv",
    primaryClass = "hep-ex",
    month = "10",
    year = "2023"
}

@article{Feng:2022inv,
    author = "Feng, Jonathan L. and others",
    title = "{The Forward Physics Facility at the High-Luminosity LHC}",
    eprint = "2203.05090",
    archivePrefix = "arXiv",
    primaryClass = "hep-ex",
    reportNumber = "UCI-TR-2022-01, CERN-PBC-Notes-2022-001, INT-PUB-22-006, BONN-TH-2022-04, FERMILAB-PUB-22-094-ND-SCD-T",
    doi = "10.1088/1361-6471/ac865e",
    journal = "J. Phys. G",
    volume = "50",
    number = "3",
    pages = "030501",
    year = "2023"
}

@article{Kling:2020mch,
    author = "Kling, Felix and Trojanowski, Sebastian",
    title = "{Looking forward to test the KOTO anomaly with FASER}",
    eprint = "2006.10630",
    archivePrefix = "arXiv",
    primaryClass = "hep-ph",
    doi = "10.1103/PhysRevD.102.015032",
    journal = "Phys. Rev. D",
    volume = "102",
    number = "1",
    pages = "015032",
    year = "2020"
}

@inproceedings{Craik:2022riw,
    author = "Craik, Daniel and Ilten, Phil and Johnson, Daniel and Williams, Mike",
    title = "{LHCb future dark-sector sensitivity projections for Snowmass 2021}",
    booktitle = "{Snowmass 2021}",
    eprint = "2203.07048",
    archivePrefix = "arXiv",
    primaryClass = "hep-ph",
    month = "3",
    year = "2022"
}

@article{LHCB-FIGURE-2025-001,
      collaboration = "LHCb",
      title         = "{Performance of LHCb as a feebly interacting particles
                       detector}",
      year          = "2025",
      url           = "https://cds.cern.ch/record/2923551",
}

@techreport{Jerhot:2936260,
      author        = "Jerhot, Jan",
      collaboration = "NA62",
      title         = "{Projection of finalized NA62 beam-dump analyses to the
                       full 2021-2026 statistics}",
      institution   = "CERN",
      reportNumber  = "NA62-PUB-2025-01",
      address       = "Geneva",
      year          = "2025",
      url           = "https://cds.cern.ch/record/2936260",
      doi           = "10.17181/CERN.ZLV5.XN5S",
}

@article{CODEX-b:2019jve,
    author = "Aielli, Giulio and others",
    collaboration = "CODEX-b",
    title = "{Expression of interest for the CODEX-b detector}",
    eprint = "1911.00481",
    archivePrefix = "arXiv",
    primaryClass = "hep-ex",
    doi = "10.1140/epjc/s10052-020-08711-3",
    journal = "Eur. Phys. J. C",
    volume = "80",
    number = "12",
    pages = "1177",
    year = "2020"
}

@article{Aielli:2022awh,
    author = "Aielli, Giulio and others",
    title = "{The Road Ahead for CODEX-b}",
    eprint = "2203.07316",
    archivePrefix = "arXiv",
    primaryClass = "hep-ex",
    month = "3",
    year = "2022"
}

@article{CODEX-b:2025rck,
    author = "Aielli, Giulio and others",
    collaboration = "CODEX-b",
    title = "{CODEX-b: Opening New Windows to the Long-Lived Particle Frontier at the LHC}: {ESPP Update Comprehensive summary}",
    eprint = "2505.05952",
    archivePrefix = "arXiv",
    primaryClass = "hep-ex",
    month = "5",
    year = "2025"
}

@article{BaBar:2015jvu,
    author = "Lees, J. P. and others",
    collaboration = "BaBar",
    title = "{Search for Long-Lived Particles in $e^+e^-$ Collisions}",
    eprint = "1502.02580",
    archivePrefix = "arXiv",
    primaryClass = "hep-ex",
    reportNumber = "BABAR-PUB-14-010, SLAC-PUB-16226",
    doi = "10.1103/PhysRevLett.114.171801",
    journal = "Phys. Rev. Lett.",
    volume = "114",
    number = "17",
    pages = "171801",
    year = "2015"
}

@article{Evans:2020aqs,
    author = "Evans, Jared A. and Gandrakota, Abhijith and Knapen, Simon and Routray, Hardik",
    title = "{Searching for exotic $B$ -meson decays enabled by the CMS L1 track trigger}",
    eprint = "2008.06918",
    archivePrefix = "arXiv",
    primaryClass = "hep-ph",
    doi = "10.1103/PhysRevD.103.015026",
    journal = "Phys. Rev. D",
    volume = "103",
    number = "1",
    pages = "015026",
    year = "2021"
}

@article{Canepa:2022swf,
    author = "Canepa, Anadi and D'Onofrio, Monica",
    title = "{Future accelerator projects: new physics at the energy frontier}",
    eprint = "2306.12897",
    archivePrefix = "arXiv",
    primaryClass = "physics.acc-ph",
    reportNumber = "FERMILAB-PUB-22-248-PPD",
    doi = "10.3389/fphy.2022.916078",
    journal = "Front. in Phys.",
    volume = "10",
    pages = "916078",
    year = "2022"
}

@article{Frugiuele:2018coc,
    author = "Frugiuele, Claudia and Fuchs, Elina and Perez, Gilad and Schlaffer, Matthias",
    title = "{Relaxion and light (pseudo)scalars at the HL-LHC and lepton colliders}",
    eprint = "1807.10842",
    archivePrefix = "arXiv",
    primaryClass = "hep-ph",
    doi = "10.1007/JHEP10(2018)151",
    journal = "JHEP",
    volume = "10",
    pages = "151",
    year = "2018"
}

@article{DOnofrio:2015gop,
    author = "D'Onofrio, Michela and Rummukainen, Kari",
    title = "{Standard model cross-over on the lattice}",
    eprint = "1508.07161",
    archivePrefix = "arXiv",
    primaryClass = "hep-ph",
    reportNumber = "HIP-2015-30-TH",
    doi = "10.1103/PhysRevD.93.025003",
    journal = "Phys. Rev. D",
    volume = "93",
    number = "2",
    pages = "025003",
    year = "2016"
}

@article{Inomata:2018htm,
    author = "Inomata, Keisuke and Kawasaki, Masahiro and Kusenko, Alexander and Yang, Louis",
    title = "{Big Bang Nucleosynthesis Constraint on Baryonic Isocurvature Perturbations}",
    eprint = "1806.00123",
    archivePrefix = "arXiv",
    primaryClass = "astro-ph.CO",
    reportNumber = "IMPU-18-0101, IMPU 18-0101",
    doi = "10.1088/1475-7516/2018/12/003",
    journal = "JCAP",
    volume = "12",
    pages = "003",
    year = "2018"
}

@article{Bagherian:2025puf,
    author = "Bagherian, Hengameh and Ekhterachian, Majid and Stelzl, Stefan",
    title = "{The bearable inhomogeneity of the baryon asymmetry}",
    eprint = "2505.15904",
    archivePrefix = "arXiv",
    primaryClass = "hep-ph",
    doi = "10.1007/JHEP01(2026)068",
    journal = "JHEP",
    volume = "01",
    pages = "068",
    year = "2026"
}
\end{document}